# Tectonics controls hydration-induced rheological heterogeneities in ultraslow-spread oceanic lithospheres

**Authors**

Leila Mezri*[1], Alexander Diehl[1], Thomas P. Ferrand[2], Javier Garcia-Pintado[1], Manon Bickert[3] and Marta Perez-Gussinye[1]

**Affiliations**

(1) MARUM - Center for Marine Environmental Sciences, University of Bremen, Bremen, Germany
(2) Institut Langevin, ESPCI Paris, Université PSL, CNRS UMR 7587, Paris, France
(3) Geo-Ocean, Univ Brest, CNRS, IFREMER, UMR6538, F-29280 Plouzané, France
(*) Corresponding author. Email : lmezri@marum.de

**Abstract**

At ultraslow, magma-poor spreading ridges, plate divergence is controlled by tectonics, with formation of detachment faults that expose variably serpentinized mantle rocks. Seismicity and the depth of seismic events in these environments are influenced by the degree of serpentinization, yet the nature and extent of deeper alteration remain poorly understood. This study uses a 2D visco-elasto-plastic model with thermodynamic calculations to explore the interaction between tectonics and hydration during ultraslow seafloor spreading. By coupling water availability and lithosphere hydration progress with active deformation, we reveal: (i) a tectonically controlled vertical extent of alteration along detachment faults; (ii) the preservation of amphibole-facies in exhumed serpentinized footwalls; (iii) significant lithospheric-scale rheological heterogeneities due to variations in alteration mineral assemblages. We propose that tectonically controlled spatial variations in alteration play a key role in controlling earthquake depth distribution at mid-ocean ridges and associated transform faults and discuss the implications for seismogenesis in subduction zones.

**Teaser**

Spatial variations in alteration influence earthquake depth distribution in ultraslow-spread oceanic lithospheres.

**MAIN TEXT**

**Introduction**

At ultraslow (< 20 mm.yr$^{-1}$), magma-poor mid-ocean ridges (MORs), newly formed oceanic lithosphere exhibits significant heterogeneities due to tectonic, magmatic, and hydration processes (e.g., (*1*)). Seismicity in these regions shows a distinctive depth distribution: sparse seismicity in shallow levels, 4-6 km depth at MORs and 5-16 km at oceanic transform faults (OTFs), overlies clusters of deeper microseismic events (8-15 km at MORs, 10-34 km at OTFs) (*2-4*). The reduced seismicity in the shallow lithosphere is attributed to serpentinization (e.g., (*3*, *4*)), which forms low-friction mineral assemblages that weaken the lithospheric mantle (*5*, *6*). This weakening potentially results in aseismic deformation (*7*). Although a strong link between serpentinization and seismogenesis in oceanic lithosphere exists (e.g., (*8-11*)), the role of the distribution of alteration assemblages at temperatures beyond the serpentine stability field remains unclear.

Using a 2D visco-elasto-plastic model (Rift2Ridge, (*12*)), with thermodynamic calculations (Perple_X, (*13*)) we simulated lithospheric hydration in ultraslow, magma-poor detachment-dominated environments. Previous versions of the code (without thermodynamic coupling) addressed tectonic, sedimentary, magmatic, and surface processes during continental rifting (*14-17*) and seafloor spreading at ultraslow magma-poor ridges (*12*). In this study, we additionally coupled water availability and rock hydration to active deformation via an empirical parameterization (see Methods). This parameterisation simulates the availability of water along fractures and faults (e.g., (*18-20*)). Contrary, absence of deformation serves as a reaction inhibitor in our parameterisation.

We find that the interaction between faulting, exhumation, and hydration produces highly heterogeneous alteration patterns related to the cyclic nature of the accretional style at ultra-slow spreading ridges, commonly known as flip-flop accretion mode (*21*, *22*). Amphibole-bearing assemblages form near the root of detachments in the high-temperature realm of the lower brittle lithosphere, whereas the shallower and colder parts are extensively serpentinized. The coupling of water availability (and hydration reactions) with active deformation results in an inhomogeneous lithosphere architecture, with the formation of intensely serpentinized domains during hydration along detachments, while other exhumed, water-limited amphibole-bearing domains within the footwall remain only partially hydrated from an earlier detachment phase during the flip-flop mode of lithosphere accretion. These amphibole-bearing mantle domains form km-scale asperity-like features. The large-scale heterogeneity of lithosphere alteration predicted by our models, results in substantial rheological changes across and between alteration assemblages due to friction reduction induced by the formation of hydrous minerals (*5*, *6*). The largest rheological changes occur along the deep hydration front near the brittle-ductile transition zone (BDT). By comparing our model results with seismic data from two magma-poor segments—the easternmost Southwest Indian Ridge (SWIR) and the Knipovich Ridge (KR) (Fig. 1)—we observe that sparsely seismically active regions correlate with highly serpentinized domains in the shallow lithosphere, while deeper seismically active zones correspond to low alteration degrees and the presence of amphibole, talc and chlorite. These findings support a conceptual model where tectonically-controlled variations in alteration assemblages induce rheological heterogeneities within the brittle lithosphere. Theses alteration assemblages influence microearthquake distributions at active MORs and OTFs and have implications for the seismogenesis within subducting oceanic lithosphere.

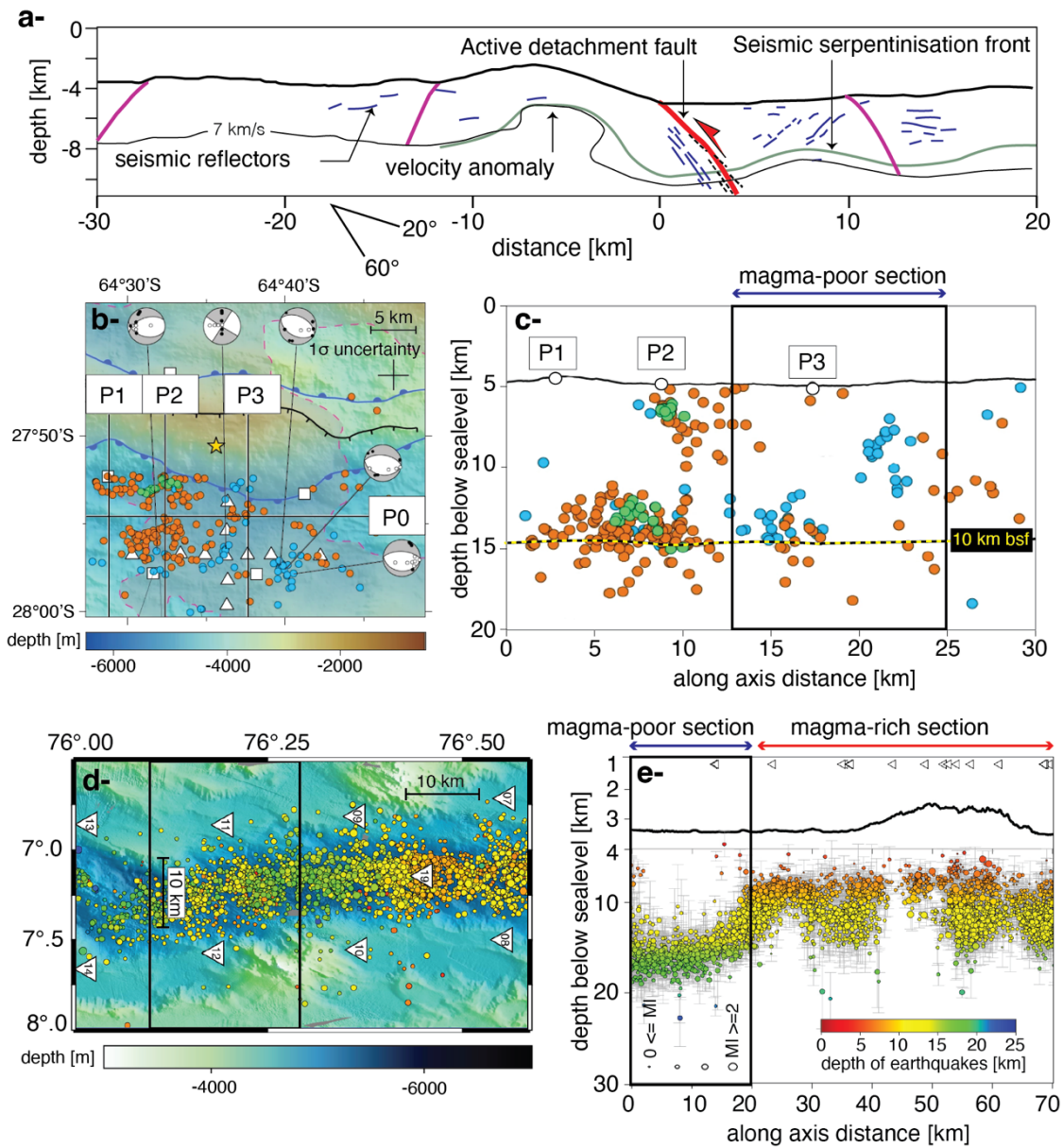

Figure 1: (a) Bathymetry redrawn from profile along the NS profile in the eastern Southwest Indian Ridge (SWIR) at 64° 30` E (modified after (*12*)), showing the 7 km.s$^{-1}$ velocity contour (black line; (*23*)); seismically detected serpentinization front (green line; (*23*)); inferred inactive detachment faults (pink lines; (*22*)); the seismically imaged active detachment (red line; (*2*)); seismic reflectors (blue lines) interpreted either as tectonic damage zones, intrusive magmatic bodies or contrasts in the degree of serpentinization of the ultramafic basement (e.g., (*24*)). (b) Bathymetric map of the SWIR 64° 30` E area and earthquake epicenters (circles) (modified from (*2*)). Blue and orange circles show seismic events of the SMSMO and RVSMO catalogs, respectively, and green circles show a seismic swarm event (*2*). The 1σ uncertainty show the average absolute horizontal location uncertainty of 3.2 km (*2*). (c) Along-axis depth profile P0' projection earthquakes within ± 8 km off the profile (modified from (*2*)). The average vertical uncertainty is ± 2.8 km. (d) Bathymetry of a section of the Knipovich Ridge with a the magma-poor section denoted in the black rectangle (modified from (*3*)). Circles show the earthquake epicenters. The color code indicates the depth of seismic events and circle size the magnitude. The white triangles show the ocean bottom seismometer (OBS) stations. (e) Along axis profile with projected earthquake positions (the gray bars indicate errors of depth estimates).

## Results

**Thermodynamic modelling of hydration**

We calculated a thermodynamic pressure-temperature (P-T) equilibrium phase diagram for the mineral composition of water-saturated abyssal peridotites in the NCFMASH system ($Na_2O$-$CaO$-$FeO$-$MgO$-$Al_2O_3$-$SiO_2$-$H_2O$) (see Methods). We assume fluid accessibility upon faulting (e.g., ([19](), [20](), [25]())), and alteration by hydration in a closed system under water saturated conditions (e.g., ([26]())). The composition of the chemical system remains unchanged except for addition of $H_2O$ according to water saturation at thermodynamic equilibrium. The software used to perform the equilibrium calculations was Perple_X ([13]()), and the solid–solution thermodynamic models were those reported in ([10]()) (see Supplementary Materials Table S3). For our reference case, the easternmost SWIR at 64° 30` E (hereafter referred to as SWIR case), we used a composite harzburgite-lherzolite composition compiled from a suite of mantle lherzolites and harzburgites recovered along the SWIR ([27]()). In the following, we refer to this composite as SWIR mantle composition (see Supplementary Materials Table S2). The KR is geochemically distinct from the easternmost SWIR. To account for the regional variations in mantle composition, we used a bulk harzburgite composition compiled from rock samples obtained along the sparsely magmatic zone in Gakkel Ridge (GR) as representative for the magma-poor section in KR ([26](), [28]()). We chose the GR rock samples, given their proximity and, because no bulk composition is available for abyssal peridotite from the KR. In the following, this composition is referred to as KR mantle composition (Supplementary Materials Table S2).

Figure 2a-b shows the P-T phase diagram of mineral assemblages for the SWIR and KR mantle compositions and predicts the P-T stability fields of the hydrous minerals, such as amphibole (amph), chlorite (chl), talc, serpentine (serp) and brucite (br). These minerals are observed in abyssal peridotites worldwide, including those from the eastern SWIR and the sparsely magmatic zone in GR ([26](), [29-31]()). For our SWIR mantle composition, our thermodynamic models predicts that amphibole is stable at $T \approx 475\text{-}750$ °C, over the full pressure range of 1.0-20.0 kbar. At $P \approx 1.0\text{-}15.0$ kbar, amphibole+talc+chlorite form an equilibrium assemblage at $T \approx 500\text{-}670$ °C and amphibole+talc+chlorite+serpentine form an equilibrium assemblage at $T \approx 500\text{-}620$ °C (Fig. 2a). At temperatures between 475 °C and 500 °C, talc is not part of the equilibrium assemblage and below 475 °C amphibole is not thermodynamically stable. At $T \leq 400$ °C, serpentinization is associated with the formation of brucite, resulting in a serpentine+chlorite+brucite paragenesis (Fig. 2a). For our KR mantle composition (Fig. 2b), our pseudosections predict a similar equilibrium assemblage as for SWIR mantle (Fig. 2a).

Figure 2c-d shows the variations in mineral fractions along a linear P-T trajectory from 1 kbar and 200 °C to 3 kbar and 750 °C. Notably this linear P-T trajectory does not exactly reproduce the geotherm at MORs, but exemplifies the evolution of stable alteration assemblages during a realistic exhumation path under thermodynamic equilibrium. For both the SWIR and KR mantle compositions, the degree of alteration, i.e., the cumulative abundance of hydrous minerals in the primary mantle assemblage, increases with decreasing temperature along the P-T path (Fig. 2c-d). According to our thermodynamic simulations, the increase in hydration degree for SWIR mantle composition is overall similar to KR mantle composition, and is associated with the following evolution of alteration mineral assemblages with decreasing temperature (Fig. 2c-d): (1) amphibole+chlorite for $T \approx 670\text{-}700$ °C; (2) amphibole+chlorite+talc for $T \approx 580\text{-}670$ °C (B); (3) amphibole+chlorite+talc+serpentine for $T \approx 580\text{-}610$ °C; (4) amphibole+chlorite+serpentine for $T \approx 550\text{-}580$ °C; (5) chlorite + serpentine for $T \approx 420\text{-}550$ °C; (6) chlorite+serpentine+brucite at temperatures

< 420 °C. Based on this evolution, we identified the following three main alteration assemblages (Fig. 2c-d): amphibole-bearing assemblage (A) from 500 °C to 700 °C, high-temperature (HT) serpentine-bearing assemblage (B) from 350 °C to 500 °C, and low-temperature (LT) serpentine-bearing assemblage (C) from 350 °C to near surface temperature conditions. The transition between the assemblages (A) and (B), which occurs between 500 °C and 550 °C, is associated with the appearance of serpentine and disappearance of talc and amphibole.

The transitions between the parageneses, from (A) to (B) and (B) to (C), are associated with an abrupt change in the degree of alteration (see red curve Fig. 2c-d). It is important to note that the above depicted evolution of mineral parageneses reflects a first-order and simplified alteration sequence expected during the retrograde exhumation of lithospheric mantle under equilibrium conditions. In nature different evolutions of mineral assemblages may occur according to tectonic context, magma supply, and chemical composition of the protolith and alteration fluids. Nevertheless, the similarities in the alteration assemblages between the SWIR and the KR mantle (with the exception of the modal mineral proportions that differ slightly between the two), suggest that the two mantle compositions in both contexts show an overall similar evolution of hydrous mineral assemblages when being exposed to similar P-T paths during mantle exhumation and hydration.

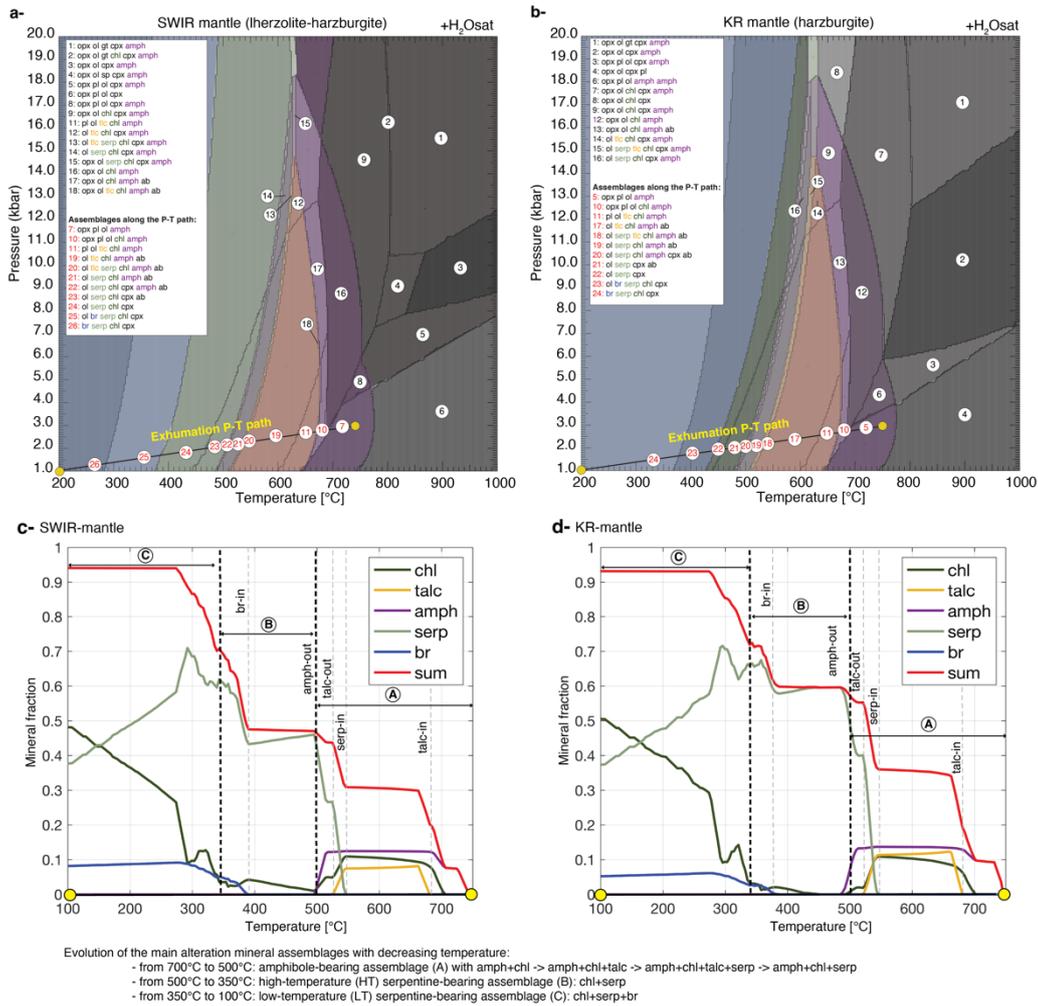

Figure 2: (a-b) Pressure-temperature (P-T) mineral phase diagrams, in the NCFMASH ($Na_2O$-$CaO$-$FeO$-$MgO$-$Al_2O_3$-$SiO_2$-$H_2O$) closed system at $H_2O$ saturated conditions for (a) Knipovich Ridge, KR, and (b) Southeast Indian Ridge, SWIR, at 64° 30` E mantle compositions (Supplementary Tab. S2), calculated using Perple_X software (*13*). The SWIR mantle composition represents a composite harzburgite-lherzolite composition compiled from a suite of mantle lherzolite and harzburgite compositions along the southwest Indian ridge (*27*). The KR mantle composition represents bulk harzburgite composition from rock samples along the sparsely magmatic zone in Gakkel Ridge (*26*, *28*). (c-d) Variations in mineral fractions and alteration degree (the cumulative abundance of hydrous minerals in the assemblages; red curve) along the linear P-T trajectories depicted by yellow circles and the straight line in panels a-b. Here we focus on the following hydrous minerals: amphibole (amph), chlorite (chl), talc, serpentine (serp) and brucite (br). Note: The yellow circles mark conditions of 750 °C and 3.0 kbar and 200 °C and 1.0 kbar.

## Faulting mode, water availability and rock hydration

At magma-poor ridges, external seawater-derived fluids penetrate the lithosphere along faults and fractures, providing pathways for serpentinization and other alteration reactions (e.g., (*18-20*)). Here, geodynamic simulations of lithospheric hydration were carried out using the version of the *Rift2Ridge* 2-D visco-elasto-plastic code presented by Mezri et al. (*12*), which we have coupled to thermodynamic calculations (see Methods) and a parameterisation of water availability as a function of strain rate. Our thermodynamic calculations are combined with a simple reaction restraint, which is the availability of water.

Thermodynamic coupling with the geodynamic simulation is achieved by incorporating the results of pre-calculated thermodynamic phase diagrams into the geodynamic model (see Methods). The pre-calculated thermodynamic phase diagrams are those for the SWIR and KR mantle compositions presented above (Fig. 2). The thermodynamic data are integrated into the geodynamic code in the form of lookup-tables, which contain the mass fractions of selected hydrous minerals (amphibole, chlorite, talc, serpentine and brucite; Fig. 2) and the water content in the water-saturated bulk rock (Supplementary Materials Fig. S2) as a function of pressure and temperature, for each mantle composition (SWIR and KR mantle composition; Supplementary Materials Table S3).

The geodynamic code solves the momentum, mass and energy conservation equations in a Lagrangian framework (see Methods). Plastic strain weakening on faults is simulated by decreasing rock cohesion and friction coefficient with increasing accumulated plastic strain until a threshold criterion is met (e.g., (*12, 14, 32*); see Methods). The effect of hydrous mineral formation with low internal friction is incorporated in the model by reducing the friction coefficient with increasing alteration degree (i.e., the cumulative abundance of hydrous mineral in the assemblage; see Methods). Viscous strain weakening along ductile shear zones is simulated by linearly increasing the pre-exponential dislocation or diffusion creep factor until a threshold is reached ((*14, 16*); see Methods).

To assess alteration architecture, we examine the dynamic evolution of faults and resulting changes in fault pattern predicted by our reference model, the SWIR case. In this model, we set the lithosphere-asthenosphere boundary temperature to 1250 °C (e.g., (*33*)) and apply half extension velocities of 15 mm.yr$^{-1}$ at the lateral boundaries (e.g., (*34*)) (Supplementary Materials Fig. S1). Figure 3a-d shows the deformation pattern (depicted by the accumulated plastic strain) and the active deformation over a 0.6 Myr spreading period starting at 21.0 Myr in the SWIR case (see also the evolution of the accumulated plastic strain and plastic/viscous active deformation in Movies S1 and S2, respectively). Our model reproduces a brittle lithosphere of ~10 km thickness (Fig. 3a-c), consistent with the depth below seafloor (bsf) at which most microseismicity occurs in the easternmost SWIR 64° 30' E segment (Fig. 1c; (*2*)). This thick brittle lithosphere is a result of the combined effects of cold mantle temperature and hydrothermal cooling along active faults (see Methods; (*12*)). At 21.0 Myr of extension, plate divergence is accommodated by two oppositely dipping conjugate normal faults, forming a horst system (cf. F1 and F2 in Fig. 3a). Subsequently, one of the two conjugate faults outcompetes its counterpart (e.g., F1) and becomes a new detachment fault, DF1 (Fig. 3b). During the early detachment phase, increased bending stresses in the rotating footwall lead to the formation of secondary antithetic and synthetic faults in both the foot- and hanging wall (Fig. 3b); as previously described by Lavier et al. (*32*). Plastic strain mainly accumulates along DF1 detachment and in the antithetic fault, F2 (Fig. 3b).

During the late detachment phase, at 21.6 Myr of extension, detachment fault DF1 accommodates plate divergence with a total displacement of up 12 km (Fig. 3c). During this late phase, the footwall domain widens, and a series of secondary faults form in the footwall due to stresses generated during flexural bending. Finally, a new antithetic fault, F3, dissects the footwall (Fig. 3c). The repeated formation of opposing detachments in the footwall of their predecessors—known as the flip-flop detachment mode (*21, 22, 35*), (Movies S1 and S2)—alternates with near-symmetric horst and spider deformation modes (*12, 36*). The models also show sequentially active synthetic accommodating faults, as observed in Chen et al., (*37*) and modelled by Glink et al., (*38*). At 21.6 Myr of extension, this leads to a seafloor morphology shaped by shallow block-crests and deep valleys, where sediment ponds form on the aging seafloor. The modeled smooth seafloor morphology reproduces the observed seafloor morphology and bathymetry from the easternmost SWIR 64° 30' E segment (compare Figs. 3c and 1a), with crests and valleys spaced by inactive detachment and dominant antithetic faults (shown by the accumulated plastic strain) (Fig. 3a-c).

Retrograde hydration reactions, including serpentinization, occur upon cooling when water is available (e.g., (*39, 40*)). The availability of water depends on rock permeability (e.g., (*41*)), which is strongly influenced by faults and fractures (e.g., (*19, 42, 43*)). Here, we evaluate the availability of water through a dimensionless parameter, $p_w$ (see Methods). This parameter is expressed as a function of an available water source term ($r_w$) and the plastic strain rate, which is used to proxy the combined effects of permeability (e.g., (*42*)) and porosity on the local availability of water to react with the model rock (Methods). The dependency of $p_w$ on plastic strain rate ensures available water is supplied only to highly localized strain faults, consistent with petrological and geophysical investigations (e.g., (*10, 19, 44*)) and experimental petrology (*45-47*). Hydration is assumed to be instantaneous, and occurs if available water is equal or larger than the mass fraction of water predicted in the thermodynamically stable, water-saturated alteration assemblages for given P-T (see Methods and Supplementary Materials Fig. S2).

To evaluate the role of faults on hydration over the 0.6 Myr period of spreading for the SWIR reference case, we focus on the spatial distribution of available water along active faults where deformation occurs (Fig. 3d-f and Supplementary Movie S3). During the horst phase, at 21.0 Myr, faults F1 and F2, which form the horst, constitute the main pathways for water from the surface to the brittle-ductile transition zone (BDT) at ~10 km bsf (Fig. 3d). Within the shallow part of the horst (T < 500 °C) the absence of active faults results in a lack of water availability (Fig. 3d). Conversely, in the lower part of the horst (T > 500 °C) active secondary faults provide sufficient water for hydration. The lower strain rates on these secondary faults result in less water availability compared to the dominant F1 and F2 faults (Fig. 3d). During the early detachment phase, the DF1 detachment and newly formed antithetic and synthetic secondary faults establish a new network of water pathways (Fig. 3e). In the last detachment phase (Fig. 3f), a new series of antithetic and synthetic secondary faults creates another network of water availability. Concurrently, the availability of water along older, long-lived secondary faults such as F2 diminishes as strain rate decreases (e.g., 21.3 Myr) and these faults progressively become inactive (Fig. 3d-f and compare Movies S2 and S3 time steps 21-21.6 Myr).

Finally, these results suggest that when water availability is coupled with plastic strain rate, changes in faulting modes during seafloor spreading, along with variations in strain rate intensity across fault segments, lead to spatial and temporal fluctuations in water availability for hydration.

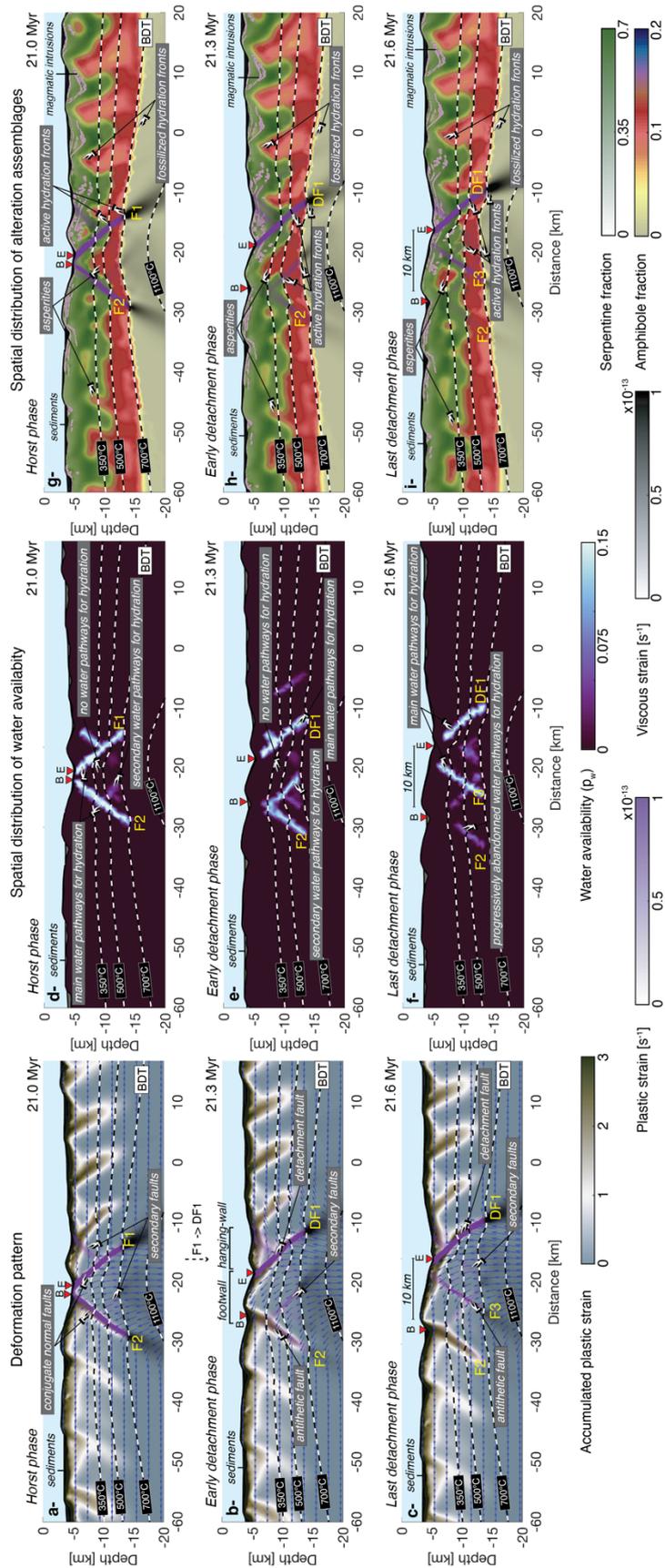

Figure 3: Dynamic evolution of faulting, water availability and hydration for the SWIR case (reference model) over a 0.6 Myr spreading period starting at 21.0 Myr. Panels a-c show the fault patterns illustrated by the accumulated plastic strain and the mantle flow velocity illustrated by the blue arrows. Panels d-f show the spatial distribution of water availability and resulting current hydration of the lithosphere ($p_w$; Methods). Panels g-i show the spatial distribution of serpentine- and amphibole-bearing mineral assemblages in altered mantle rocks. In panel i, zones 1, 2 and 3 correspond to the lithospheric regions between surface conditions to the 350 °C isotherm, between 350 °C and 500 °C isotherms, and between 500 °C and 700 °C isotherms, respectively. In zone 1 the dominant alteration assemblages is brucite+chlorite+serpentine, chlorite+serpentine in zone 2 and amphibole+talc+chlorite in zone 3. Active plastic and viscous strain are shown in purple and black transparency scales. The dashed black and white lines show the following isotherms: 350 °C, 500 °C (approximation of the serpentinization front) and 700 °C (approximation of the deepest hydration fronts and the BDT). The letters E and B, represent detachment emergences and footwall breakaway zone, respectively. Red triangles indicate the position of E and B. The full model evolution, from continental break-up to 21.6 Myr, for the accumulated plastic strain (a-c), water availability (d-f) and lithosphere hydration (g-i) can be seen in Supplementary Movies S1, S3 and S4, respectively.

**Alteration architecture and associated mineralization**

In this study, we simulate lithosphere hydration during tectonically dominated, magma-poor seafloor spreading, focusing on the alteration assemblages predicted for specific mantle compositions (see Fig. 2a-b, Supplementary Table S2). In our simulations, alteration of the model rocks by hydration is restricted to active brittle faults, where water supply is sufficient (Fig. 3d-f; see Methods). We analyze this process in the reference model (Southwest Indian Ridge, SWIR case) by examining the evolution of alteration assemblages with decreasing temperature on active dominant faults (F1, F2, DF1, and DF3; see Fig. 3d-f). We compare these alteration assemblages with the across-axis alteration patterns in the SWIR and Knipovich Ridge (KR) models (Figs. 3g-i, 5; Movie S4).

In the SWIR case, along dominant active faults, strain-dependent water availability drives hydration, leading to the formation of amphibole-bearing assemblage between 500 °C and 700 °C. Serpentinization occurs between 500 °C and near surface temperature conditions (Fig. 4). This evolution is consistent with petrological observations in samples from the SWIR 64° 30 `E segment (*29*) and Gakkel Ridge (*26*). Amphibole, chlorite, and talc appear when peridotites are incipiently hydrothermally altered around 670 °C and 700 °C. In our model this temperature range marks a transition from unaltered mantle to amphibole-bearing assemblages, which we call the deep hydration front (around the BDT). During hydration between 500 °C and 550 °C, serpentine forms, and amphibole and talc are replaced at T < 500 °C. Consequently, we consider the 500 °C isotherm as the high-temperature (HT) serpentinization front, which separates amphibole-free from amphibole-bearing assemblages. At temperatures below 350 °C, brucite is part of the equilibrium assemblage containing serpentine and chlorite. We call this transition the low-temperature (LT) serpentinization front (Fig. 4).

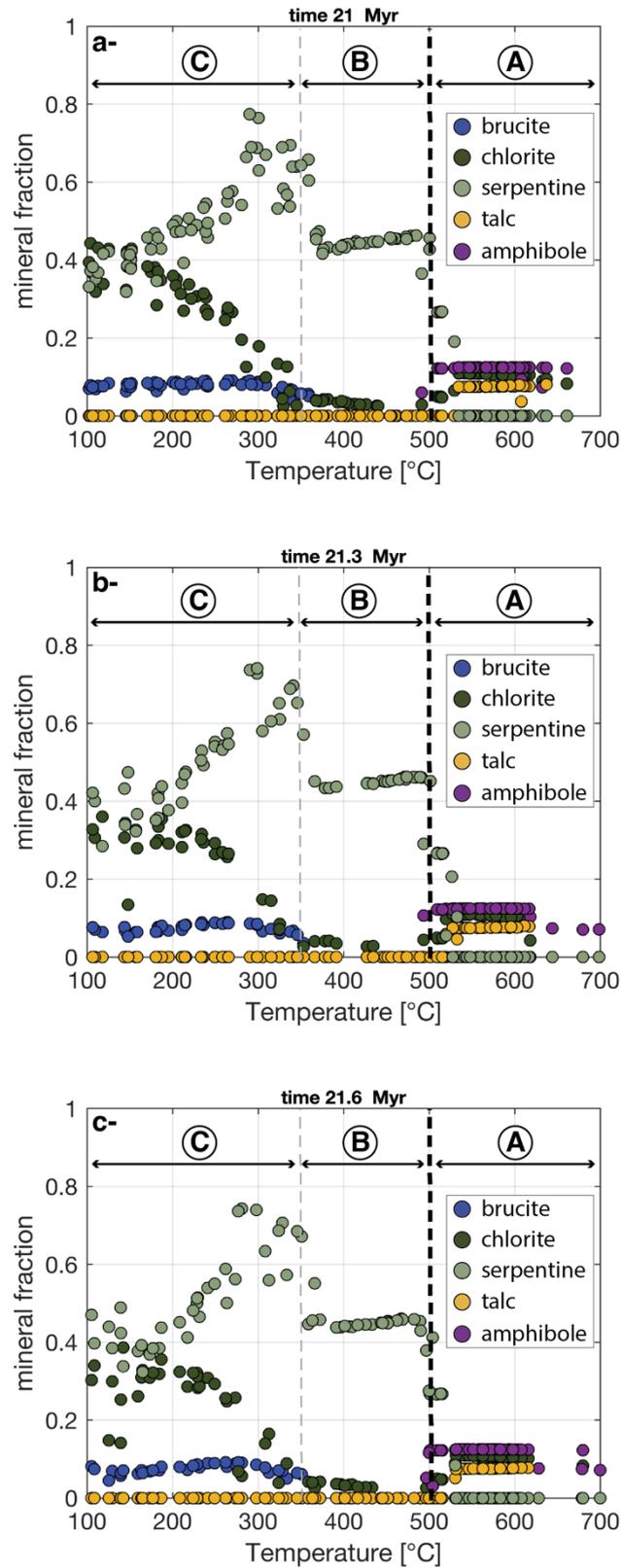

Figure 4: Panels a-c show the evolution of the alteration assemblage with decreasing temperature along dominant active faults in the reference model, the SWIR case, for each time steps shown in Figure 3 (see caption Figure 2 for details on the model). The letters A, B and C corresponds to the following assemblages: amphibole-bearing assemblage, high- and low-temperature serpentine-bearing assemblage, respectively. These assemblages are consistent with those predicted by our thermodynamic calculation (see Fig. 2c-d).

During the horst phase in the SWIR case (at 21.0 Myr), hydration occurs along the active faults where water is available, but not within the shallow part of the horst, due to the absence of active deformation and water availability (compare Figs. 3d and 3g). As a result, amphibole-bearing alteration assemblages are exhumed without proceeding to hydrate, although entering the thermodynamic stability field of serpentinization under low-temperature conditions (T < 500 °C) (Fig. 3g). Consequently, the water-limited amphibole-bearing mantle domain preserves its high temperature phase-assemblage despite entering the serpentine stability field and forms an asperity-like structure (Fig. 3g).

During the early detachment phase at 21.3 Myr, the amphibole-bearing mantle domain continues to be exhumed along detachment fault DF1 without being overprinted by secondary low-temperature mineral assemblages in the footwall (Fig. 3h, Movie S4 time steps 21-21.3 Myr). Contrary, hydration occurs in the lower part of the footwall, along DF1 and secondary faults, driven by strain-dependent water availability (compare Figs. 3e and 3h, and Movies S2, S3 and S4, time steps 21.0-21.3 Myr). Variations in strain rate, and consequently water supply (which depends on the strain rate; see Methods), yield spatial variations in mineral proportions across alteration assemblages (Figs. 3i and 5, and compare Movies S2 and S4).

As detachment fault DF1 evolves, new antithetic and secondary faults form, further hydrating the footwall and transforming locally the metastable amphibole-bearing assemblage into serpentine-bearing assemblage (Fig. 3h-i and Movie S4 time step 21.3-21.6 Myr). Along the detachment and dominant antithetic faults, strain-dependent hydrothermal cooling enhances hydration to reach greater depths than within the footwall. Along the active detachment, the LT serpentinization front at 350 °C, the HT serpentinization front at 500 °C, and the deep hydration front at 700 °C extend to depths of 3-4 km, 6-7 km and 10-11 km bsf, respectively. The deepening along detachment faults aligns with both seismic observations (compare Figs. 3i and 1a), and previous numerical modeling studies (*12, 23*).

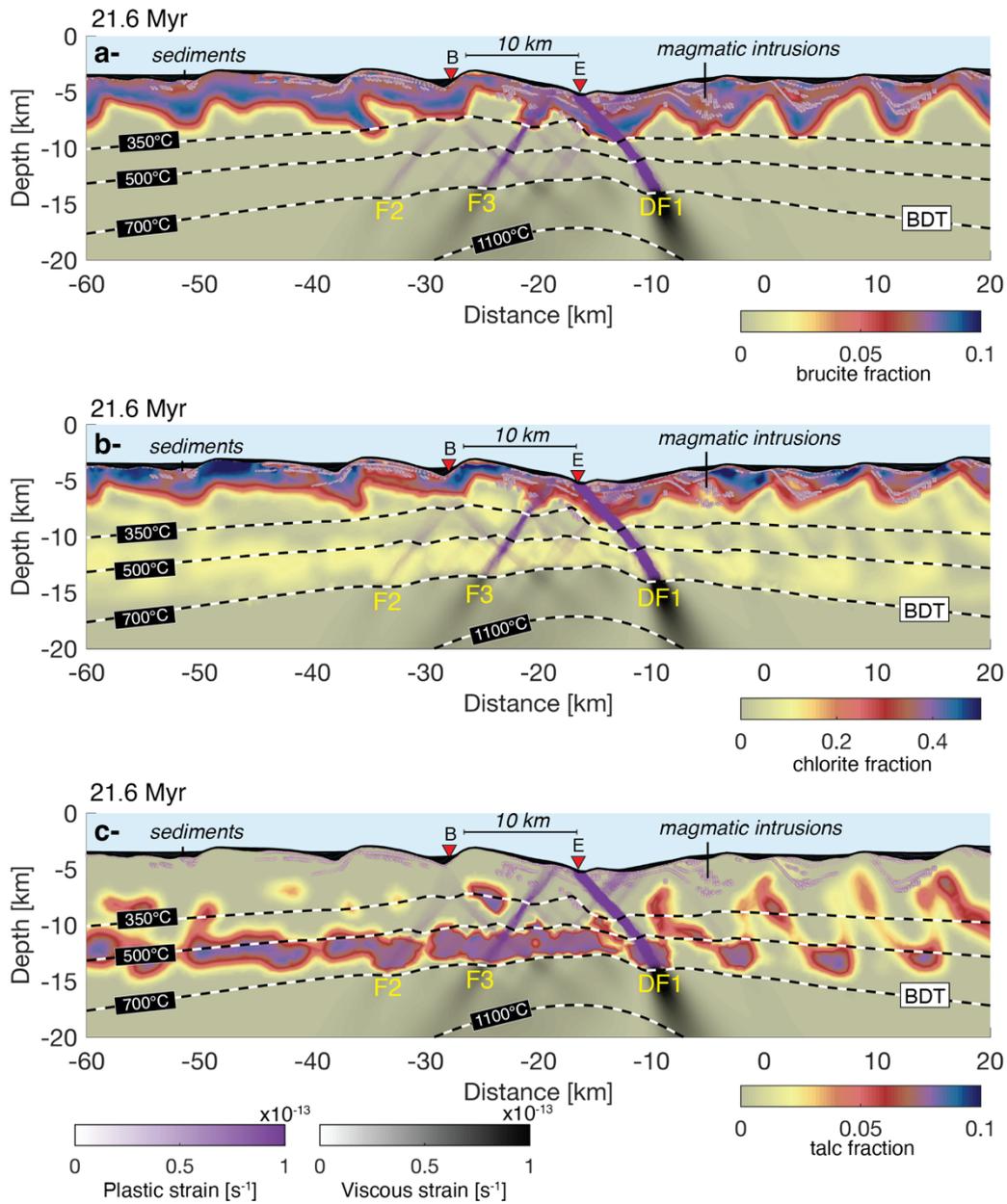

Figure 5: Spatial distribution of brucite (a), chlorite (b) and talc (c) in the reference model SWIR case after 26.1 Myr of extension. In all panels a-c, active plastic and viscous strain are shown in purple and black transparency scales. The dashed black and white lines show the following isotherms: 350 °C, 500 °C (approximation of the serpentinization front) and 700 °C (approximation of the deepest hydration fronts and the ductile brittle transition zone, BDT). The letters E and B, represent detachment emergences and footwall breakaway zone, respectively. Red triangles indicate the position of E and B.

We compare the lithosphere hydration architecture between the Knipovich Ridge (KR) and the Southwest Indian Ridge (SWIR) (compare Figs. 3i and 6a). For the KR case (Fig. 6 and Movie S5), we applied a higher asthenosphere temperature of 1310 °C (e.g., (48)), compared to 1250 °C used for the SWIR case. The absolute spreading rate along the Knipovich Ridge is about 14-16 mm.yr$^{-1}$ (e.g., (49)), similar to the eastern SWIR (34). However, the Knipovich

Ridge is highly oblique, with an angle of 40°-53° between the ridge and the spreading direction, which influences faulting mode (*49*). In contrast, the angle between the ridge and the spreading direction along the eastern segment at 64°30`E in SWIR is about 33° (*34*). To account for the effect of the strong obliquity along the KR, we applied an effective lower spreading rate of 10 mm.yr$^{-1}$, whereas a rate of 15 mm.yr$^{-1}$ is applied for the SWIR case. For the KR case, hydration is simulated using a harzburgite mantle composition (KR mantle composition; Figs. 2b and 2d and Supplementary Materials Table S2).

Comparing the SWIR and KR cases (Figs. 3i and 6a) reveals that, despite differences in spreading rate, mantle composition, and mantle temperatures, the faulting mode and alteration architecture are similar. Both cases (SWIR and KR cases) exhibit a shallow serpentinized mantle domain transected by metastable amphibole-bearing assemblage forming asperities, overlying a lower brittle lithosphere made up of amphibole-bearing assemblage (Movie S5). As the lithosphere ages, both inactive fault zones (depicted by the accumulated plastic strain; Fig. 3a-c and Movie S1) and the alteration architecture formed at the active ridge become fossilized (Fig. 3g-i and Movies S4 and S5). This inherited alteration architecture remains preserved as long as no active deformation occurs to supply water for further hydration overprinting.

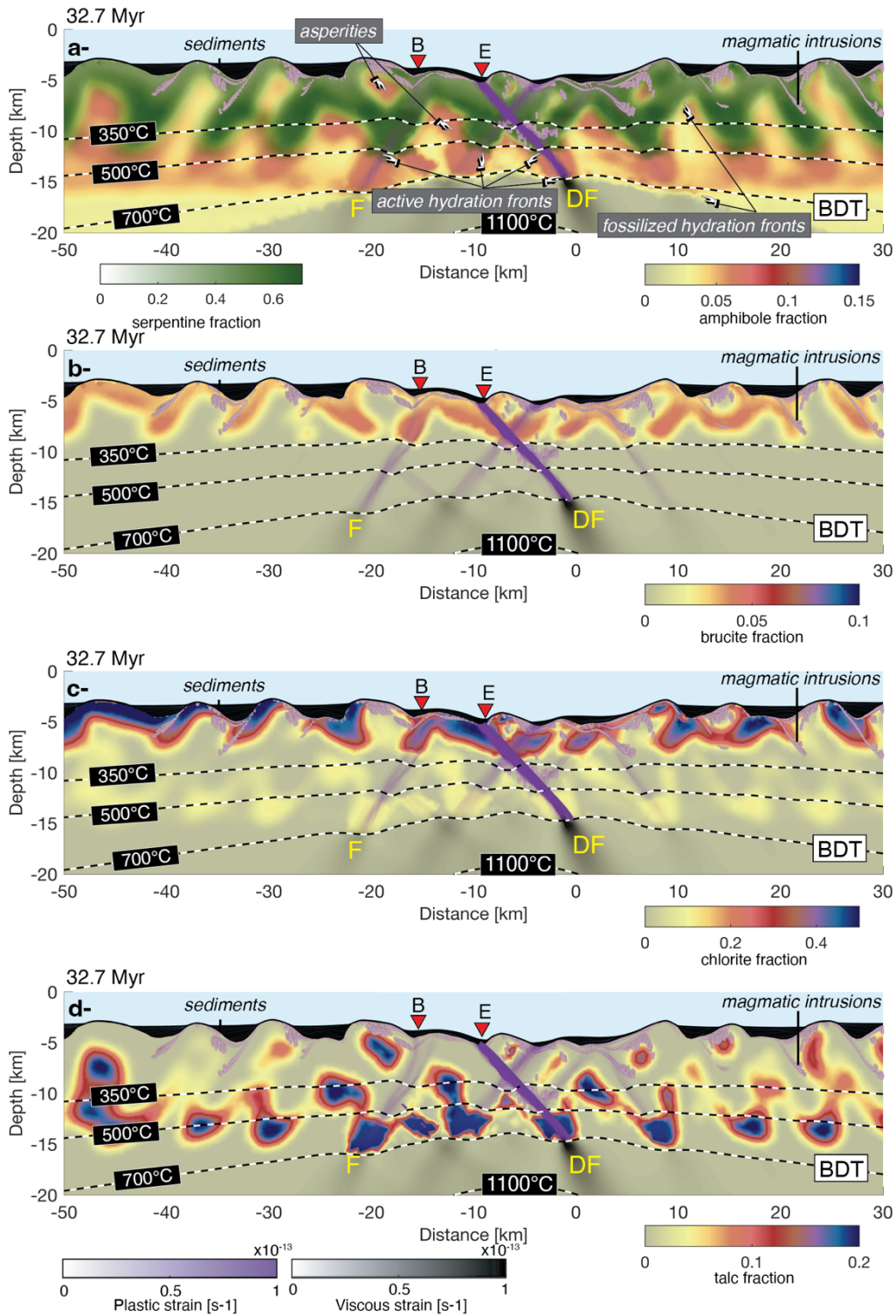

Figure 6: Spatial distribution of alteration mineral assemblages in the oceanic lithosphere for the KR case after 32.7 Myr of extension. Panel a shows the spatial distribution of serpentine- and amphibole-bearing mineral assemblages in altered mantle rocks. Spatial distribution of brucite (a), chlorite (b) and talc (c) in the reference model. In all panels a-d, active plastic and viscous strain are shown in purple and black transparency scales. The dashed black and white lines show the following isotherms: 350 °C, 500 °C (approximation of the serpentinization

front) and 700 °C (approximation of the deepest hydration fronts and the ductile brittle transition zone, BDT). The letters E and B, represent detachment emergences and footwall breakaway zone, respectively. Red triangles indicate the position of E and B.

In our model, we use the magma emplacement procedure described in (*12*) (Methods). At each time step (10 kyrs), the volume of melt produced beneath the ridge is assumed to ascend "instantaneously" and settle as a dike at shallow depths (Methods). The vertically rising magma stalls either just below the top of the basement or at greater depth if it encounters highly serpentinized mantle rocks (> 70 %), which we assume to act as a barrier to magma ascent in the context of a low magma budget (e.g., see (*12, 16*)). Shallow magma emplacement mainly occurs during the horst phase, due to relatively high temperatures within and low degree of serpentinization (see (*12*)). During the detachment phase, however, the deepening serpentinization front along the detachment results in progressively deeper magma emplacement in the footwall (e.g., (*12*)).

Once emplaced, the magma releases latent and sensible heat, causing a localized temperature increase of up to 100 °C in the surrounding area (Methods; (*12*)). This localized heating both limits hydration and locally weakens the rocks promoting strain localization (see (*12*)). Secondary faults, that form in the footwall and hanging wall, cut through the already accreted magma, allowing serpentinization below the magma emplacement depth (*12*). Retrograde hydration reactions are exothermic (e.g. (*50*)) and here, for the sake of simplicity, we only consider the exothermic heat of serpentinization (Methods; (*12*)), which causes a temperature increase of less than 50 °C in regions undergoing serpentinization. While this temperature increase would delay hydration (*50*), it has no significant effect on the model's timescale due to hydrothermal cooling along the faults where serpentinization occurs.

**Implications for depth-distribution of earthquakes**

Changes in mantle peridotite rheological properties arise from the localization of deformation and weakening processes, such as grain size reduction (*51, 52*), fluid-assisted mechanisms like pressure–dissolution processes (*53*), pinning in polyphase rocks (*54*), and the formation of hydrous mineral with low internal friction (*53*). In our model, mantle weakening occurs as friction reduces with increased accumulated plastic strain on faults and alteration degree—the cumulative abundance of hydrous minerals (see Methods; red curve in Fig. 2c, d). We assume the overall mantle friction coefficient is governed by the weaker of the two effects—either alteration or accumulated plastic strain. The minimum value between the two determines the mantle's friction coefficient (see Methods). We assess the effects of plastic strain and alteration on mantle rheology for both SWIR and KR by isolating each factor's impact on friction (Fig. 7). The effect on the mantle's rheology is evaluated by calculating the depth derivative of the friction coefficient.

In our model, the friction coefficient reduces as the accumulated plastic strain increases from 0 to 1, reaching minimum values after sustained deformation (> 0.5 Myr; Fig. 7a-d and Movies S1 and S2). In both the SWIR and KR cases, this weakening in turn promotes the development of long-lived detachment faults with durations over 1-1.2 Myr and offsets exceeding 10-12 km, as well as, long-lived antithetic faults that accommodate footwall rollover (see Movies S4 and S5).

Increased flexural stresses in the footwall leads to the formation of secondary faults, which form a distributed network in the deeper and hotter footwall region, accumulating minimal plastic strain with limited effect on friction (Fig. 7a-f). Simultaneously, hydration along active faults alters mantle rocks (Figs. 3i, 4, 6a). The increase in the modal proportion of hydrous minerals, i.e., the degree of alteration, within the assemblage results in a reduction of the friction coefficient (Fig. 7e-h; see Methods). The degree of alteration diminishes with increasing temperature, as HT assemblages contain lower modal proportions of hydrous phases than LT assemblages. Consequently, along transitions between alteration assemblages and around metastable HT amphibole-bearing assemblages surrounded by serpentinized mantle rocks, this result in abrupt variations in friction coefficient and substantial rheological changes (Fig. 7e-j).

Unlike the localized effect of accumulated plastic strain along dominant fault zones, spatial variations in the degree of alteration lead to highly heterogeneous friction patterns in the brittle lithosphere (Fig. 7a-h). These heterogenous friction patterns lead to contrasting rheologic behaviour, which remains fossilized in the aging lithosphere (Fig. 7g-j).

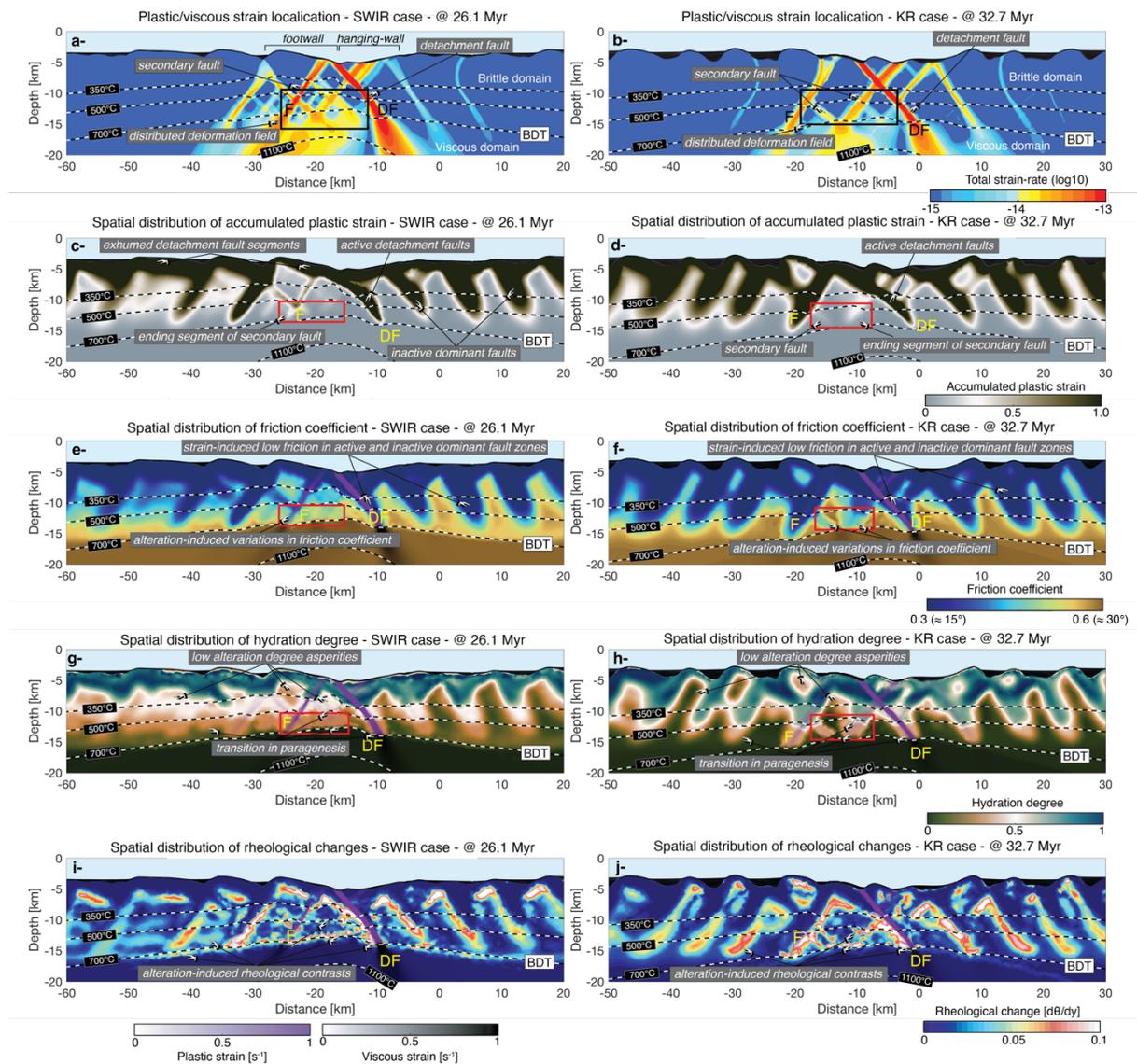

Figure 7: Panels on the left show model results for SWIR and on the right for KR cases. Panels a-b show the spatial distribution of the logarithm of the total strain rate. Panels c-d show the fault patterns depicted by the accumulated plastic strain. Panels e-f show the spatial distribution of the friction coefficient of the mantle. Panels g-h show the spatial distribution of the degree of hydration, i.e., the cumulative abundance of hydrous minerals. Panels i-j show the degree of rheological change, i.e., the depth derivative of the friction coefficient. In panels e-j, the active plastic and viscous strains are represented by purple and black transparency scales. The dashed black and white lines show the following isotherms: 350 °C, 500 °C (approximation of the serpentinization front) and 700 °C (approximation of the deepest hydration fronts and the ductile brittle transition zone, BDT).

We explored the link between alteration-induced heterogeneities and the depth-distribution of seismic events beneath ultraslow-spreading MORs. Earthquake nucleation often involves fault heterogeneities, including variations in mineral composition, frictional properties, and porosity (e.g., (*45, 55, 56*)). These heterogeneities induce spatial variations in stress and strength, with implications for the spatial distribution of seismicity (e.g., (*8, 9, 45, 57, 58*)). At the Southwest Indian Ridge (SWIR) at 64 °30 `E and the Knipovich Ridge (KR) magma-poor segment at 76 °25' N (Fig. 1), most seismic events occur between ~8-15 km depth, with shallower events occurring sparsely due to serpentinization (e.g., (*3*)) (Figs. 1, 8a, 8b). Increased deep microseismicity and regional clustering of events is interpreted to result from heterogeneities in deformation behavior as a result of differences in mineralogy and rheological properties near the BDT region (*29, 59*). Here, we hypothesize that the spatial variations in alteration degree across the brittle lithosphere modulates the distribution of seismicity. To test this hypothesis, we compare our model results with the depth distribution of seismic events from both the SWIR and KR regions (Fig. 8a-b).

For this comparison, we consider the lateral extent of earthquake data across the SWIR and the KR ridge segments, covering ~10 km and ~20 km wide zones, respectively (Figs. 1b, d; 1, 8). The depth distribution of seismic events for both regions (Fig. 8a-b) is compared with depth variations in laterally averaged mineral fractions and alteration degree (Fig. 8c-d), friction coefficient (Fig. 8e-f), and rheological changes (the depth derivative of the friction coefficient; Fig. 8g-h). For each depth, we calculate the average along the x-direction within a specified window, which aligns with the extent of the earthquake datasets from the SWIR and the KR regions. For each case, SWIR and KR, the window is centered on the spreading center in our models, where the horizontal velocity along the model topography is zero ($v_{x_{topo}} = 0$). The model's average friction coefficient ($\theta_{model}$; Fig. 8e-f) accounts for the effect of accumulated plastic strain and hydration. It is compared to the friction coefficient, which depends on the accumulated plastic strain only ($\theta_{post-proc}$ is calculated in post-processing; see gray curve Fig. 8e-f). We also compared the rheological impacts of both $\theta_{model}$ and $\theta_{post-proc}$ (see dark-blue and light-blue curves, respectively; Fig. 8g-h).

In both cases, alteration decreases with depth and increasing temperature (Fig. 8c-d). Along the 350 °C, 500 °C, and 700 °C isotherms, transitions in mineralogy between assemblages leads to a reduction in the degree of alteration and a corresponding increase in friction ($\theta_{model}$; Fig. 8c, f). The depth-amplitude of friction changes increases as the degree of alteration decreases (Fig. 8c, f). The depth-amplitude peaks along the BDT (~10-12 km depth), where fresh mantle is replaced by amphibole-bearing assemblages, reducing the friction up to 10 % ($\theta_{model}$; Fig. 8c, f). This variation, the largest and most abrupt one predicted by our model, drives the most substantial rheological changes across the entire brittle lithosphere ($\theta_{model}$; Fig. 8e-h). Conversely, when friction depends solely on accumulated plastic strain, it remains nearly constant between 500 °C and 700 °C, with minor variations along the BDT (gray vs. black curves; Fig. 8e-f). This is due to secondary faults near the BDT accumulating insufficient strain to reduce the friction coefficient significantly (Fig. 8c-f). This therefore suggest that rheological changes in the lower and hotter brittle lithosphere primarily arise from transitions in mineral equilibria within alteration assemblages.

In both the SWIR and KR regions, the reported presence of a ~4-5 km thick in-depth band of deep microseismicity—where most earthquakes occur—in ~8 km to ~15 km depth range is associated with high-temperature mantle in the BDT region (Fig. 1 and 8a-b). The observed increase in microseismicity with depth in both regions correlates with decreasing alteration degree, as well as, with an increase in the magnitude of alteration-induced rheological changes (Fig. 8). The peak in rheological change, occurring as fresh mantle is replaced by amphibole-bearing assemblages along the BDT (~10 km depth in SWIR; ~12 km depth in KR), aligns with the seismicity peak in these regions (Fig. 8a-b).

These findings therefore highlight a link between alteration-induced rheological heterogeneities and depth-distribution of earthquakes at ultraslow-spreading ridges, with the depth extent of hydration fronts playing a crucial role. The correlation between seismic peaks and rheological contrasts along deep hydration fronts along the BDT, suggests that high-temperature alteration of upwelling mantle plays a pivotal role in deep seismic clustering within the lithosphere.

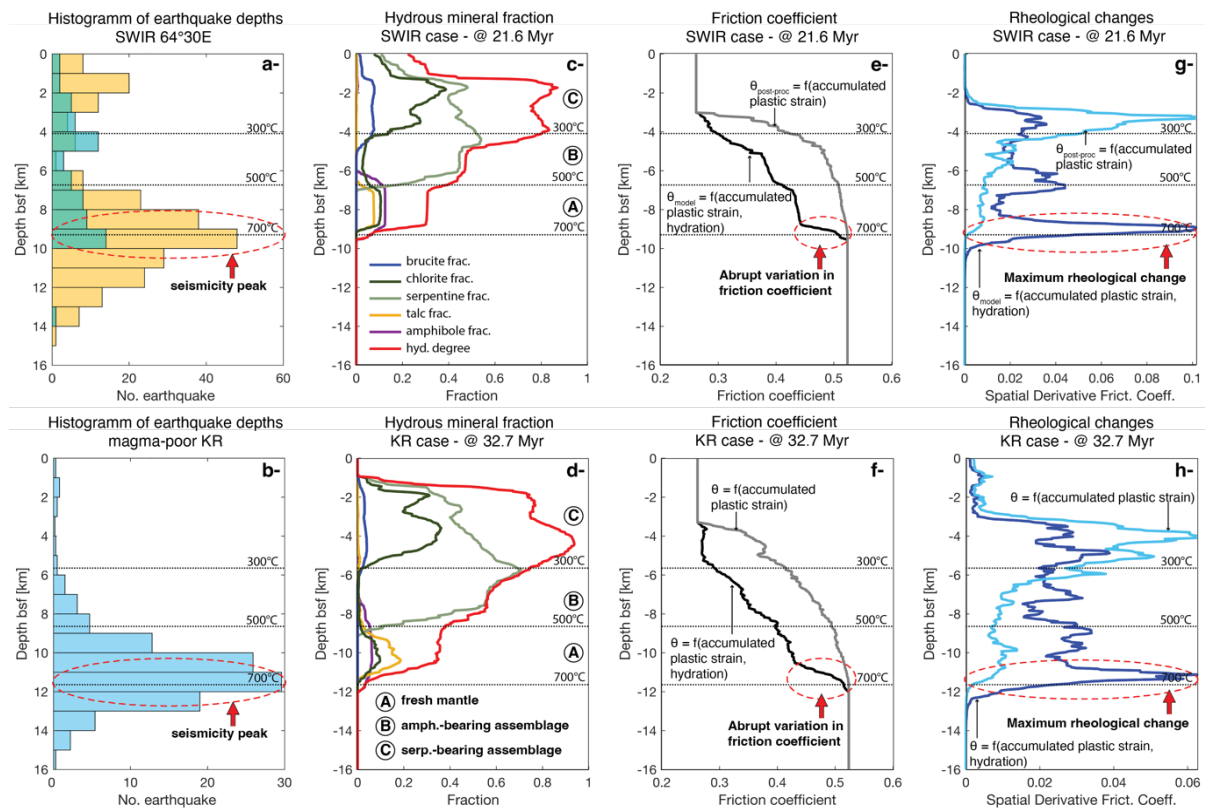

Figure 8: For SWIR (top) and KR (bottom) cases, from the left to the right, panels show depth distribution of earthquake events (earthquake histogram; g-h) at SWIR 64 °30 `E (*2*) and the magma-poor Knipovich Ridge section at ~ 76 °25 N (*3*), laterally averaged mineral fractions (c-d), friction coefficient (e-f) and depth derivative of the friction coefficient (rheological contrasts; g-h). In panels c-d, the letters A, B and C correspond to the following assemblages: amphibole-bearing assemblage, high- and low-temperature serpentine-bearing assemblages. In panel e-f, the black curve corresponds the model output's average friction coefficient, which depends on both accumulated plastic strain and hydration ($\theta_{model} = \theta = min(\theta_{hyd}, \theta_{acc})$; see Method), and the gray curve corresponds to average friction coefficient that is expressed as a function of accumulated plastic strain only ($\theta_{post-proc} = \theta_{acc}$; see Methods). In panels g-h, the dark-blue and light-blue curves show $\partial \theta_{model}/\partial y$ and

$\partial\theta_{post-proc}/\partial y$, respectively. The lateral extent of earthquake across the ridge is ~10 km at SWIR and ~40 km at Knipovich Ridge (see Fig. 1b, d). The model outputs are laterally averaged accordingly and centered at the spreading center.

## Discussion

This study provides a geodynamic view on lithosphere alteration architecture, emphasizing three critical aspects: (i) a tectonically-controlled vertical extent of alteration along detachment faults; (ii) the preservation of amphibole-facies cores in exhumed serpentinized footwalls; (iii) substantial lithosphere-scale rheological heterogeneities due to variations in alteration mineral assemblages. Before elaborating these aspects and the implication of alteration on depth-distribution of earthquake in the lithosphere at mid-ocean ridges, transform faults and subduction zones, we discuss the model mineral composition of alteration assemblages during lithospheric hydration in comparison with natural samples from the eastern SWIR at 64 °30 `E and the sparsely magmatic zone in the ultraslow Gakkel Ridge.

### Potential alteration assemblages

At slow and ultraslow magma-poor MORs, seawater-derived fluids percolating in the lithosphere alter mantle peridotites (e.g., (*1*)). Samples from the SWIR 64 °30 `E segment (~15 mm.yr$^{-1}$ full spreading) and the magma-poor section of the Gakkel Ridge (~10 mm.yr$^{-1}$ full spreading) contain similar alteration assemblages, suggesting comparable hydrothermal alteration conditions with almost no magmatic influence (*26, 30, 31*). We hypothesize that these alteration conditions are applicable to the magma-poor section of the Knipovich Ridge (KR). The seafloor of the easternmost SWIR 64 °30 `E segment displays corrugated detachment surfaces which expose abyssal peridotites with variable degrees of serpentinization (e.g., (*60*)). These samples record variable degree of alteration from serpentinization to high-temperature amphibole-facies conditions near the BDT. The observed mineralogical assemblages record a chronological sequence through exhumation, which reach from high-T, low alteration degree close to the BDT (*29*), to high alteration degree via serpentinization at lower temperatures of T < 500 °C.

Consistent with amphibole stability temperatures, recorded in samples from the SWIR 64 °30 `E segment (*29*) and Gakkel Ridge (*26*), our thermodynamic modelling of closed system hydration under saturation condition for our SWIR and KR mantle composition predicts that amphibole is stable at T ≈ 500-700 °C (Fig. 2). Our model also predicts that talc forms during hydration of peridotite at temperatures T > 500 °C and is stable until T ≈ 670 °C with amphibole and chlorite (Fig. 2). However, at the eastern SWIR, talc is nearly absent, with rare exceptions as orthopyroxene replacement, or as vein mineralization (*30*). At Gakkel Ridge, 17 % of the analysed samples show talc occurrence with amphibole and chlorite formed at high temperature conditions (*26*). The (rare) observations of talc in peridotites from magma-starved regions, such as the sparsely magmatic zone in Gakkel Ridge and the eastern SWIR, suggest that talc formation occurs only in fluids with an increased silica activity due to orthopyroxene dissolution in peridotite (*26, 30, 61-63*).

Our thermodynamic modelling predicts that brucite is stable at low temperatures T <≈ 400 °C (Fig. 2), whereas it is absent in the SWIR 64 °30 `E samples (*30, 31*). This absence is explained by either the higher concentrations of silica in the serpentinizing fluids or by subsequent dissolution and replacement of brucite upon higher water-rock ratios (*31*). In nature, hydration of rocks occurs in an open system, where fluids and rocks do chemically evolve during successive hydration (e.g., (*64*)). Consequently, pH values and silica activity of hydrothermal fluids evolve during exhumation and ongoing hydration (e.g., (*65*)). At low temperature conditions, T <≈ 400 °C, in the absence of gabbro, the silica activity would remain low, so that no low-temperature talc will form and brucite would eventually be formed and preserved at temperatures < 400 °C (e.g., (*65*)). Under low magma-budget, talc and brucite could form during incipient hydrothermal alteration at low water-rock ratios, with talc formed in small proportion at high temperatures (T = 500-670°C; Fig. 2), due to dissolution of orthopyroxene (e.g., (*26, 29, 31*)). As hydration continues, brucite would be replaced by serpentine during subsequent low-temperature fluid rock interactions that overprints early-stage serpentinization.

Therefore, we propose that beneath magma-poor ultraslow spreading ridges, an assemblage of talc+amphibole+chlorite likely forms at the root of detachments during incipient hydration, as predicted by our model. In nature, this assemblage would form during early stage of exhumation at high temperature conditions, under relatively high silica activities resulting from dissolution of orthopyroxene present in the primary mantle assemblage (i.e., (*26*)). Since talc is one of the weakest hydrous minerals (*6, 58, 66*), its formation at the base of the brittle lithosphere, even in small amounts, could significantly affect detachment localization—an effect not addressed in this study.

**Role of tectonics during lithospheric alteration**

Along ultraslow, magma-poor, spreading MORs (< 20 mm.yr$^{-1}$), plate divergence is controlled by tectonics, with part of the divergence that is accommodated by detachment faulting (e.g., (*60, 67*)). These detachment faults cut through a thick brittle lithosphere (~8-13 km; (*68*)), and are reproduced in our simulations (Figs. 3g-i, 6a, and Movies S4 and S5). We find that fault-controlled hydrothermal cooling along detachment faults increases the depth extent of hydration fronts, in agreement with geological and seismic observations (*26, 29, 60, 69*), and numerical models (*12*). However, despite geological evidence of deep hydration at temperatures greater than 500 °C to near the brittle-ductile transition (BDT; (*26, 29*)), little is known about high-temperature alteration and its spatial distribution in the lithosphere (e.g., (*60*)).

Our geodynamic model provides insights into the spatial distribution of alteration and associated mineralization at lithospheric scales (Figs. 3-6, and Movies S4 and S5). Importantly, we show that during tectonically-dominated seafloor spreading, in the context of low magma supply, the interplay between faulting, exhumation and strain-dependent water availability for hydration leads to highly heterogeneous alteration patterns (Fig. 3). Changes in faulting mode shape, both, the morphology of the seafloor and the alteration architecture of the oceanic lithosphere (see Movies S4 and S5).

During flip-flop detachment faulting at ultraslow, magma-poor MORs, detachment phases alternate with various faulting phases, such as horst phase and sequentially active detachment faults (renamed here Sequential Phase) (*12, 36, 38*). We suggest these variations play a crucial role in controlling the architecture of alteration during lithosphere accretion (Fig. 9). In the horst phase (HP; Fig. 9a), hydration occurs along the two dominant conjugate normal faults, which form the horst, and secondary active faults in flanks, as well as, the lower part of the horst (Fig. 9a). Simultaneously, the absence of faults in the shallow part of horst features limits hydration, allowing the exhumation of metastable amphibole-bearing assemblages to shallow depths (< 4 km bsf) and low temperatures (T < 500 °C) (Movie S4 time steps 20-21 Myr).

During the early detachment phase (DP-1; Fig. 9b), the amphibole-bearing assemblages form at high-temperature conditions and continue to be exhumed without being hydrated. Concurrently, surrounding rocks undergo serpentinization along active faults and fractures (Fig. 9b). As the detachment slips (DP-2; Fig. 9b and Movie S4 time steps 21-21.3 Myr), flexural bending of the footwall results in a series of antithetic and synthetic secondary faults dissecting the area (e.g., (*12, 32, 36, 70*)). These faults create new fluid pathways for hydration (Movies S2 and S3) while hydrothermal cooling extends the depth of hydration fronts along these faults (Fig. 9c and Movie S4 time steps 21-21.3 Myr). Consequently, newly formed secondary faults in the active footwall promote further hydration, reshaping locally the alteration architecture in the footwall (Fig. 9b-c and Movie S4 time steps 21-21.3 Myr). The new detachment then separates the formerly active footwall into two sections, converting a part of the former footwall into the hanging-wall of the new detachment (Fig. 9b). The abandoned detachment system, including its foot- and hanging-wall, moves passively away from the spreading center, preserving its alteration architecture, as long as no new faults form to induce additional hydration overprint (Fig. 9c).

During the sequential phase (SP; Fig. 9c), the same scenario as described during DP occurs on each of the sequentially active detachments. As a result, along both detachments, metastable amphibole-bearing assemblages are exhumed and preserved in the serpentine-stability field, in their respective footwall. Simultaneously, serpentinization occurs along the two detachments and secondary faults. When one of the two detachments accommodates the divergence, the abandoned detachment system preserves its alteration architecture (Fig. 9c).

The recurrence of this combination of processes, involving faulting, exhumation and water-limited hydration allows high- and low-temperature alteration assemblages to coexist at the same depths (P-T conditions) depending on their formation history in the accretion zone. This results in a highly heterogeneous alteration pattern in, both, the young and old oceanic lithosphere (Figs. 3-6 and 9d). Heterogeneous alteration patterns are characterized by lateral variations in the depth extent of hydration fronts, spatial variations in mineral fractions within stability fields, and the coexistence of alteration assemblages forming ~4-5 km wide asperities in the shallower region (Fig. 9d and Movies S4-S5). These asperities consist of metastable amphibole-bearing assemblages preserved in the serpentinized mantle domains, at temperatures < 500 °C.

In contrast to the shallow lithosphere, where large-scale, metastable, zone-forming asperities transect the serpentinized mantle, our model predicts that the lower part of the lithosphere is almost entirely altered (Fig. 9e and Movies S4 and S5). This alteration occurs due to the more distributed nature of deformation at the base of the brittle lithosphere under high temperature condition, T > 500 °C (Fig. 7a-b). Consequently, as the exhumed fresh mantle crosses the deep hydration front along the BDT, it inevitably passes through this distributed deformation field, promoting widespread hydration at high-temperature conditions (Fig. 9; compare Movies S2 and S4). Within this field, the intensity of the strain rates varies significantly (Fig. 7a-b). The dependency between water availability for hydration and strain rate (see Methods), leads to spatial differences in the proportion of hydrous minerals in the amphibole-bearing assemblage (Fig. 3i, 4-6 and Movies S4 and S5).

Finally, we propose that the interaction between faulting, exhumation, and hydration during flip-flop lithosphere accretion is responsible for highly heterogeneous alteration patterns related to the cyclic nature of this accretion style. These tectonically-induced alteration heterogeneities then fossilize in the aging lithosphere. This highlights the key role of tectonics in alteration architecture in young and old ultraslow spreading oceanic lithosphere (Fig. 9d).

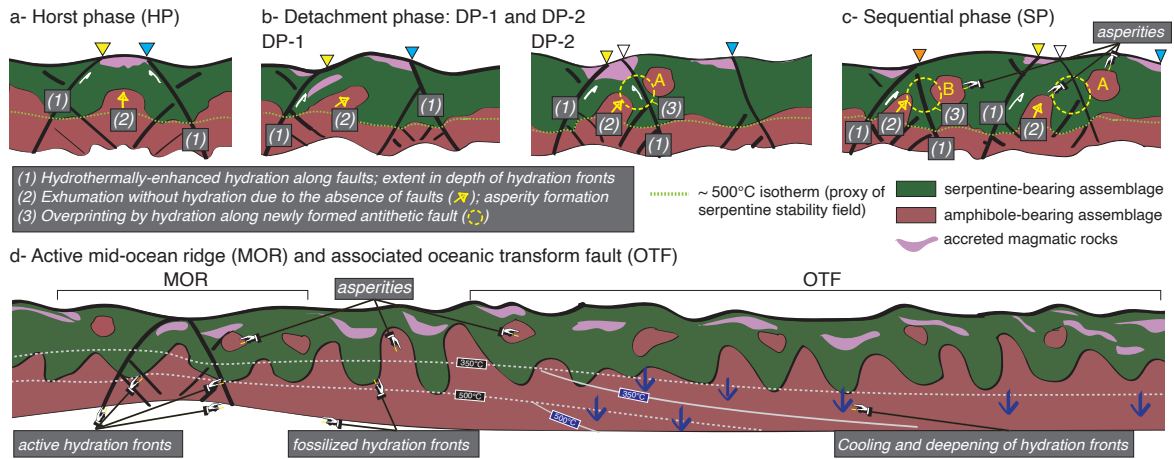

Figure 9: (a-c) Sketch evolution showing the role of tectonics on alteration architecture and associated assemblages during magma-poor, ultraslow, lithosphere accretion. The triangles indicate the dominant faults, with changes in triangle colors providing a visual chronology of fault development. The letters A and B represent the material and the numbers (1), (2) and (3) correspond to tectono-metamorphic processes. During the horst phase (HP, a) two conjugate normal faults (yellow and blue triangles) accommodate the divergence. From (a) to (b), one of the two conjugate faults (yellow triangle) becomes the new detachment (b). During the detachment phase (DP, b), a new antithetic fault (white triangle) forms to accommodate footwall rollover. During the sequential phase (SP, c), both the prior detachment (yellow triangle) and a newly formed detachment (orange triangle) simultaneously accommodate the divergence. (d) Large-scale across-axis profile illustrating alteration during accretion of new lithosphere at the active mid-ocean ridge (MOR), fossilized alteration patterns in the aging lithosphere, and cooling and hydration fronts deepening along the oceanic transform fault (OTF).

**Link between alteration and seismogenesis**

Beneath ultraslow, magma-poor mid-ocean ridges (MORs), the seismicity distribution reveals a distinct pattern: a sparse occurrence of seismic events in the upper lithosphere contrasts commonly occurring earthquakes at deeper levels below the serpentinized mantle domain (T > 500 °C; (*3*)). For example, at the eastern SWIR 64 °30 `E and the Knipovich Ridge (KR) segments, shallow seismic events are barely observed between ~4 km and ~6 km depth, while earthquake clusters are typically observed at greater depths, ranging from ~8 km to ~15 km (*2, 3*).

Microearthquake studies from active oceanic detachment faults in the North Atlantic Ocean have similarly shown seismic activity to be scarce, at depths lower than 4 km bsf (*71, 72*). The reduction of shallow seismic activity is attributed to aseismic slip in serpentinized mantle rocks (e.g., (*3, 71, 73*)). Our model reveals that in the shallow region of the oceanic lithosphere, at < 4 km depth bsf, rocks are most deformed and altered, with serpentinization degrees ranging from 50 % to 80 % (Fig. 3i, 6a, 7e f and 8c d, and Movie S7). This indicates that at these depths, the rheological rock properties are largely governed by the weaker alteration minerals, rather than by the otherwise competent primary mantle modal composition (e.g., (*6, 66, 74*)). In combination with the accumulation of significant plastic deformation, this leads to the weakest rheology among the entire brittle lithosphere. Ultimately, slip along shallow fault segments, including detachments, in this part of the lithosphere is likely accommodated aseismically (e.g., (*56*, *75*, *76*)), explaining the reduced seismicity in this region (e.g., (*3*, *71*, *73*)). Although most seismological studies indicate sparse seismic activity in the shallow lithosphere, Craig and Parnell (*77*) show that slip on detachment faults within the upper few kilometers is capable of producing large-magnitude earthquakes, which challenge the prevailing view that the shallow hydrated lithosphere necessarily behaves aseismically.

In our model, we demonstrate that substantial mechanical contrasts arise from tectonically emplaced, discrete metastable zones with various mineralogy and rheology in this shallow, serpentinized mantle region (Figs. 3i, 5a, 7g-h, and 9). Specifically, the amphibole-bearing hydration-inhibited assemblage is significantly stronger than the surrounding serpentine-bearing assemblages. As the detachment fault slips, this stronger assemblage is exhumed along the fault within just a few kilometers (less than 2 km laterally) from the fault zone.

We propose that, during slip on the detachment, small ruptures may coalesce, both, at the interface between the amphibole-bearing metastable assemblage and the surrounding material (forming the core of the shallow detachment zone) and within the metastable assemblage itself. This feature may be strong enough to trigger large earthquakes. This suggests that large earthquakes in the shallow lithosphere—typically dominated by low-friction, serpentine-bearing assemblages—can be initiated by the presence of tectonically emplaced, strong metastable zones (Fig. 9).

However, in addition to being not commonly observed, the frequency of these large earthquake events is much lower than that of microearthquakes associated with slip on the deeper portions of detachment faults (e.g., (*77*)). In our model, these metastable zones—capable of clustering large earthquakes—are separated by wide areas of serpentinized mantle which behave aseismically. Additionally, as detachment faults evolve, hydration along newly formed secondary faults in the shallow region further transform metastable assemblages into stable, aseismic serpentine-bearing assemblages. This suggests that the involvement of metastable assemblages in seismic clustering within the shallow lithosphere is spatially and temporally localized during tectonic spreading.

In contrast to the shallow lithosphere, the deeper, brittle portion of the lithosphere, from approximately 500 °C to 700 °C and between 6 km and 11 km bsf, exhibits a more uniform alteration pattern dominated by assemblages of amphibole, talc, and chlorite (Fig. 3i, 5 and 6; Movies S4 and S5). This mineral assemblage forms under high-temperature conditions along deep secondary faults and the lower segments of detachments (Figs. 3i, 5, 6, and Movies S4 and S5) and is consistent with samples from the eastern Southwest Indian Ridge (SWIR) at 64 °30 `E and the sparsely magmatic segment at Gakkel Ridge (*26*, *29*).

Despite its apparent homogeneity, variations in strain rate in the distributed deformation field leads to lateral variations in the modal proportion of hydrous minerals within this assemblage (Figs. 5-6, and compare Movies S2 and S4). These compositional changes lead to lateral variations in the friction coefficient within the lower brittle lithosphere (Fig. 7i, j). We propose that at high temperatures and high deformation degrees, these frictional heterogeneities promote semi-brittle deformation within mylonitic shear zones, potentially triggering deep earthquakes (e.g., (*10*, *78*)).

Our model suggests that the most pronounced mechanical contrasts arise along stable and metastable high-temperature serpentinization fronts, and the deep hydration front at the brittle-ductile transition (BDT). The deep hydration front in our model is responsible for the largest rheological contrast across the entire brittle lithosphere (Figs. 7-8). We suggest that these mechanical contrasts may lead to a heterogeneous distribution of velocity-weakening (strong mineral assemblages) and velocity-strengthening (weak mineral assemblages) properties, influencing spatiotemporal variations in stress accumulation and release, both of which are critical for earthquake nucleation (e.g., (*25, 55, 79*)). We therefore propose that the depth extent of these hydration fronts in the oceanic lithosphere could play a key role in determining the depth distribution of earthquakes beneath mid-ocean ridges (Fig. 9).

The depth-distribution of seismic events along the Romanche transform fault is also located below the serpentinized mantle domain (*4*). The serpentinized mantle domain, extending from 5 to 16 km in depth, is interpreted to behave aseismically. This contrasts with the deeper mantle below, from 10 to 34 km, where cluster of deep earthquakes are observed. Moving away from the ridge axis, the deep seismic band deepens, associated with cooling of the aging lithosphere (*4*). Our model predicts that the old lithosphere preserves, both, its deformation and alteration history, as well as, the resulting rheological heterogeneities acquired during accretion (Movies S4 and S5). We propose that these inherited features, and in particular the alteration patterns, influence the seismicity distribution along oceanic transform faults and fracture zones (Fig. 9d), as suggested for instance by seismicity in the Wharton basin (*9*). The cooling by seawater circulation and advection in the aging lithosphere thickens the thermal structure along the OTF, which in turn deepens the hydration fronts (Fig. 9e). Similar to alteration-induced rheological heterogeneities that control seismicity distribution beneath ultraslow, magma-poor MORs, we suggest that the increase in depth extent of hydration fronts also explain the observed deepening of the seismicity along OTFs. Along the Romanche OTF, seismic asperities are interpreted as dry mantle peridotite (e.g., (*80*)). We propose that along OTFs, asperities may also consist of altered mantle peridotite with distinct paragenesis and alteration degree, which contrast with their surroundings (Fig. 9).

In the subducting lithosphere, a significant part of the seismicity occurs at intermediate depths (~30-300 km) along the upper and lower Wadati-Benioff planes (e.g., (*25, 81, 82*)). Although it is well established that the latter are ubiquitous features of subduction zones, their origin remains debated (e.g., (*25, 83, 84*)). Petrological and geophysical observations and numerical modeling show that the structural and compositional features fossilized in the subducting lithosphere, in particular oceanic plates formed by slow and ultra-slow accretion, significantly impact the distribution of intermediate-depth seismicity (e.g., (*25, 45, 85*); and others). We show that oceanic lithosphere formed in ultraslow, magma-poor spreading environments exhibits fossilized heterogenous alteration patterns, creating discrete mechanical boundaries, which likely impact the stress field evolution within subducting slabs (*86*). These fossilized features vary considerably in degree and nature, i.e., in alteration paragenesis, both laterally and vertically in space (Fig. 9d). When subduction initiates, the oceanic plate bends and new faults localize (e.g., (*44*)). This bending-related faulting way in trench allows percolation of cold seawater, which would enable cooling and overprinting of fossilized alteration assemblages formed during lithosphere accretion. This overprinting and cooling along these faults would contribute to the depth extent of hydration fronts and introduce new rheological contrasts locally. We suggest that within the subducting slab this variability can lead and/or contribute to spatial and temporal variations in (1) fluid transport and release mechanisms (*25, 85, 87*), (2) dehydration-induced stress transfers (*8, 81*) and (3) subsequent local mechanical instabilities (*8, 25, 88*). Together, these processes play a crucial role in generating earthquakes in subduction zones.

In conclusion, this thermodynamic and geodynamic modeling study offers new insights into the alteration processes of ultraslow-spreading oceanic lithosphere in magma-poor environments. We demonstrate that the spatial distribution of oceanic seismicity at mid-oceanic ridges correlates with rheological heterogeneities caused by variations of equilibrium alteration mineral assemblages. These heterogeneities, are potentially caused by disparate water availability in the framework of tectonic lithosphere accretion. This water availability likely governs the spatiotemporal distribution of seismogenic zones, offering an explanation for observed seismicity patterns. Finally, our integrated model suggests that variations in alteration extent controls the spatial distribution of seismicity in the oceanic lithosphere at magma-poor ridges, associated transform faults, and subduction zones.

**Methods**

  1- **Geodynamic framework**

In this study, we simulated lithosphere hydration during seafloor spreading using a 2D visco-elasto-plastic code (*12*), coupled with thermodynamic calculations. Previous versions of the code (without thermodynamic coupling) addressed tectonic, magmatic, and surface processes during continental rifting (*14-17*), and seafloor spreading at ultraslow magma-poor ridges (*12*).

The code solves the momentum, mass, and energy conservation equations through a Lagrangian framework branched from MILAMIN finite element solvers (*89*). Deformation, pressure, and temperature are calculated by solving the Stokes force-balance equation, mass conservation equation assuming incompressibility, and thermal energy conservation equation, respectively. The model time step is 10 kyr.

Deformation and pressure are calculated by solving the Stokes force-balance equation (1.1) and the equation for mass conservation (1.2) assuming incompressibility:

$$\nabla \cdot \tau - \nabla P + \rho g = 0, \quad (1.1)$$

$$\nabla \cdot v = 0, \quad (1.2)$$

where $\tau$ is the deviatoric of the Cauchy stress tensor, P is total pressure, $\rho$ is density, g is the gravitational acceleration constant, and $v$ is the velocity vector.

Temperature is calculated by solving the thermal energy conservation equation:

$$\rho C_p \cdot \frac{\partial T}{\partial t} = \nabla \cdot (k \nabla T) + Q, \quad (1.3)$$

where $C_p$ is the specific heat capacity, $T$ is temperature, $t$ is time, $k$ is thermal conductivity, and $Q$ (measured in W.m$^{-3}$) the heat sources, such that

$$Q = Q_r + Q_p + Q_s + Q_{ml} + Q_{ms}, \quad (1.4)$$

where $Q_r$ is radioactive heating, $Q_p$ is shear heating by plastic deformation (depending on stress and plastic strain rate), $Q_s$ is heat release by the exothermic serpentinization reaction, and $Q_{ml}$ and $Q_{ms}$ are, respectively, the latent and sensible heat release by melt emplacement. The Stokes and mass balance equations (1.1) and (1.2) are solved together and are followed by the solution of the heat conduction equation (1.3). In addition to equation (1.3), the thermal energy is advected with the Lagrangian moving mesh.

The initial model geometry is 400 km wide and 150 km thick. It includes a continental lithosphere made of three layers, which represents the upper and lower continental crust, and the mantle (Supplementary Fig. S1a). We use wet quartzite and wet anorthite rheologies for the upper and lower continental crust, respectively (Supplementary Fig. S1c). For the mantle, including the lithosphere and asthenosphere separated by a 5 km transition zone, we use a dry and wet olivine rheology, respectively (Supplementary Fig. S1c). In the model, specific thermomechanical properties are applied for each layer (Supplementary Table S1).

The model domain is discretized using a triangular mesh (Supplementary Fig. S1a-c), employing second-order Lagrange elements for temperature, discontinuous linear elements for pressure, and Crouzeix-Raviart elements for velocities (*90*). Along the x-direction, the spatial resolution is set to 5000 m at the base of the mantle, 4000 m at the continental crust/mantle boundary, 2000 m at the upper/lower crust boundary, and 1000 m at the surface. The resolution is increased using a mesh refinement procedure, doubling the resolution, over a 200 km wide and 60 km thick region at the center of the domain from the top (Supplementary Fig. S1a, b, d).

## 2- Rheology

The visco-elasto-plastic rheology is obtained by an additive decomposition of the deviatoric strain-rate into elastic, viscous and plastic components. The constitutive relation between shear stress $\tau$ and strain rate, $\dot{\varepsilon}$, is:

$$\tau = \eta_{eff}\left(2\dot{\varepsilon}' + \frac{\tau^{oldJ}}{\mu \Delta t}\right), \qquad (2.1)$$

where $\eta_{eff}$ is the effective viscosity, $\dot{\varepsilon}'$ is the deviatoric strain rate, $\tau^{oldJ}$ is the Jaumann corotational stress, $\mu$ is the shear modulus and $\Delta t$ is the time step (10 kyr). The effective viscosity depends on the deformation regime; namely whether the material undergoes plastic yielding or deforms in the visco-elastic regime. To account for this, we employ a Drucker-Prager yield criterion for plasticity

$$\sigma_{yield} = P\sin\phi + C\cos\phi, \qquad (2.2)$$

where $\sigma_{yield}$ is the yield stress, $P$ is pressure, $C$ is the cohesion of the rocks, and $\phi$ is the friction angle. The material deforms plastically when $\tau_{II} \geq \sigma_{yield}$, where the subindex II indicates the square root of the second invariant, and $\tau_{II}$ refers to the deviatoric stress.

To include plasticity into the viscous formulation, we use a Prandtl-Reuss flow law, so that the effective viscosity $\eta_{eff}$ of a yielding material (e.g., (*91*)) is defined as

$$\eta_{eff} = \frac{\sigma_{yield}}{\left(2\dot{\varepsilon}' + \frac{\tau^{oldJ}}{\mu \Delta t}\right)_{II}}. \qquad (2.3)$$

When $\tau_{II} < \sigma_{yield}$, the material behaves visco-elastically, and the effective viscosity is

$$\frac{1}{\eta_{eff}} = \left(\frac{1}{\eta_{dis}} + \frac{1}{\eta_{dif}} + \frac{1}{\mu\Delta t}\right), \qquad (2.4)$$

where $\eta_{dis}$ is the dislocation creep viscosity, and $\eta_{dif}$ is the diffusion creep viscosity, such that

$$\eta_{dis/dif} = F\, B^{-\frac{1}{n}} \dot{\varepsilon}_{II}^{\frac{1-n}{n}} exp\left(\frac{E + PV}{nRT}\right), \qquad (2.5)$$

where $F$ is a factor for scaling parameters obtained from uniaxial/triaxial experiments to the second invariant-based formulation of the viscosity used here (*14*), $B$ is the dislocation of diffusion creep constant, n is a power-law exponent, $E$ is the activation energy, $V$ is the activation volume, $R$ is the gas constant, $T$ is the absolute temperature, and $\dot{\varepsilon}_{II}$ is the square root of the second invariant of the deviatoric strain rate.

The effect of mantle depletion resulting from melting is taken into account in the calculation of the density and viscosity. The density is calculated following the Boussinesq approximation as:

$$\rho = \rho_0(1 - \alpha(T - T_0) - \beta F_m), \qquad (2.6)$$

where $\rho$ is the density, $\alpha$ is the thermal expansivity, $\beta$ ($\beta = 0.044$ (*92*)) is a factor that parametrizes the influence of melt extraction on density, $F_m$ is the mantle depletion, $\rho_0$ is the reference density (Supplementary Tab. S1), and $T_0$ is a reference temperature. Continental lithospheric mantle is simulated as dry olivine, and the asthenosphere as wet olivine (~500 ppm H/Si (*93*)). As melting occurs, the viscosity linearly changes from wet olivine to a final dry olivine after a depletion of 4 %.

### 3- Model components

The model includes Winkler bottom condition (*94*), a dynamic topography calculated by using a stress-free surface (*14, 15*), and mantle melting and melt emplacement with associated heat released, ocean loading, and strain-rate dependent hydrothermal cooling (*12*). Hardening mechanisms are not included.

The melt production in our model is as implemented in (*12*). Following (*95*), partial melting of dry mantle occurs when the temperature exceeds the pressure-dependent and depletion-dependent solidus temperature, which is given by

$$T_s = T_s^0 + \frac{\partial T_s}{\partial P}P + \frac{\partial T_s}{\partial F}F_m, \qquad (3.1)$$

where $T_s^0$ is the solidus temperature at the surface, $\frac{\partial T_s}{\partial P} = 132$ °C /GPa is the solidus-pressure dependence, $\frac{\partial T_s}{\partial F} = 350$ °C is the solidus-depletion dependence, P is the pressure and $F_m \in [0, 1]$ is the depletion; the ratio between extracted melt and solid residue (*95*).

Here, we adjusted the surface anhydrous solidus temperature to align with the specific mantle rock types examined in our study. For the Southwest Indian Ridge (SWIR) mantle, consisting of composite harzburgite-lherzolite composition (Supplementary Table S2), we used a solidus temperature of 1147 °C (*96*). For the Knipovich Ridge (KR) mantle, consisting mainly of harzburgite composition (Supplementary Tab. S2), we used a solidus temperature of 1114 °C, consistently with harzburgite solidus temperatures reported in other studies (*95, 96*). During melting, the temperature of the mantle decreases to the solidus temperature due to the consumption of the latent heat of fusion. As a result, the geothermal gradient is reduced in this region.

At each time step, 80 % of the total volume of melt produced is assumed to be instantaneously extracted and transported to the surface (*12, 97*). We assume that the ascending melt travels vertically from the shallowest point of the 1100 °C isotherm, the starting point of the melt's ascent. Thus, the dike is emplaced at shallow depths at the horizontal location determines by the shallowest point of the 1100 °C isotherm, which follows the overlying deformation. Along its vertical route, the magma emplaces below the surface or below the serpentinization front, assuming serpentinization acts as a barrier for melt ascent (*12*). The magma body, i.e., dike, is then emplaced with a temperature of 1100 °C and the latent and sensible heat released by crystallization is homogeneously distributed throughout this time interval (*12*).

The model includes a parameterized hydrothermal cooling that consist of multiplying a nominal thermal conductivity by a so-called Nusselt number ($Nu$) (e.g., (*98*)). Cooling occurs along active brittle faults (e.g., (*42*)), which serve as proxies for water availability, as faults and fractures constitute efficient fluid conduits (e.g., (*19, 43*)). We achieve this by setting Nu as a function of a nominal plastic strain rate and a corresponding nominal $Nu_{max}$. The deformation-dependent function to estimate $Nu$ is given by

$$Nu = 1 + (Nu_{\max} - 1)\left(1 - e^{\left(-\frac{3\dot{\varepsilon}_{II}^{p}}{\dot{\varepsilon}^{H}}\right)}\right), \qquad (3.2)$$

where $\dot{\varepsilon}_{II}^{p}$ is the second invariant of the plastic strain-rate, which is used as a proxy for permeability. In this study, we set $Nu_{\max} = 14$, and $\dot{\varepsilon}^{H} = 1 \times 10^{-13}$ s$^{-1}$ (*12*). Thus, the intensity of cooling is modulated by the intensity of plastic strain, such that cooling on fault zone, assumed to be a consequence of cold seawater circulation along active faults, increases with increasing strain-rate and reaches a maximum intensity for $\dot{\varepsilon}^{H}$.

4- **Lithospheric hydration within a geodynamic framework**

During retrograde metamorphism, hydration, more than pressure and temperature, depend critically on the presence of water (e.g., (*40, 99*)). In our model, mantle hydration occurs along active brittle faults, use as proxies for water availability, as faults and fractures constitute efficient fluid pathways (e.g., (*19*)).

***Conditions for hydration during retrograde metamorphism***

First, we examine the pressure-temperature $(P - T)$ conditions that satisfy retrograde metamorphic conditions:

$$\begin{cases} P^t \leq P^{t-1} (1) \\ T^t \leq T^{t-1} (2) \end{cases} \quad (4.1)$$

where $P^t$ and $T^t$ are the pressure and temperature at time $t$, and $P^{t-1}$ and $T^{t-1}$ are the pressure and temperature at the previous time step $(t - 1)$.

Once conditions (1) and (2) are met (equation 4.1), we evaluate the hydration condition (3) (equation 4.2), which depends on water availability:

$$p_w \geq r_{sat}^{eq} \quad (4.2)$$

where $p_w$ represents the water availability in the system at a given time and $r_{sat}^{eq}$ is the fraction of water content (weight %) at saturation thermodynamic equilibrium of the corresponding mineral assemblage at given $P - T$ conditions (Supplementary Fig. S2).

*Water availability*

Hydration, including serpentinization, of ultramafic rocks at ridges occur upon cooling of the oceanic lithosphere when (sea-)water is available (e.g., (*18, 39, 100*)). We consider that faults and fractures are the main conduits for water to ingress the crust and mantle (e.g., (*18-20*)). Consistently with our parameterization for strain-rate dependent hydrothermal cooling (equation 3.2), at each time step and nodal position, we define the water availability ($p_w$) as follows:

$$p_w = r_w \left(1 - e^{\left(-\frac{3\dot{\varepsilon}_{II}^p}{\dot{\varepsilon}^H}\right)}\right) \quad (4.3)$$

where $\dot{\varepsilon}_{II}^p$ is the second invariant of the plastic strain-rate, $\dot{\varepsilon}^H$ is the threshold strain-rate value ($\dot{\varepsilon}^H = 5^{-14} \, s^{-1}$) and $r_w$ is the fixed source available water; $r_w = 0.15$ represents the mass fraction of water in the 30 % of porosity of a peridotite rock with an initial density of 3360 kg. m$^{-3}$ (e.g., (*101*)). The water availability ($p_w$) ranges from 0 to $r_w$, with $r_w = 0.15$ reached for the threshold strain-rate value ($\dot{\varepsilon}^H$).

*Thermodynamic and thermomechanical coupling*

The coupling between thermodynamic and geodynamic processes is achieved by incorporating pre-computed tables into our geodynamic model (e.g., (*102-105*)). These tables contain the fractions of hydrous minerals $f_i(P, T)$, which represents the fraction of hydrous mineral $i$ as a function of $P - T$, and the water content at saturation thermodynamic equilibrium conditions as a function of $P - T - r_{sat}^{eq}(P, T)$.

The pressure-temperature (P-T) phase diagrams are obtained by Gibbs energy minimization for specific chemical compositions of abyssal peridotite (Supplementary Tab. S2), in the NCFMASH (Na$_2$O-CaO-FeO-MgO-Al$_2$O$_3$-H$_2$Osat) system, assuming hydrothermal alteration in a closed chemical system (e.g., (26)). We performed all thermodynamic calculations using *Perple_X* software (13), and the solid solution thermodynamic models used are those reported in ((10); Supplementary Tab. S3). From these precalculated thermodynamic models, we extracted $r_{sat}^{eq}(P,T)$ and $f_i(P,T)$. For this study, we selected the following hydrous minerals: amphibole (amph), chlorite (chl), talc (tlc), serpentine (serp) and brucite (br).

*Model implementation*

During the model run, we use the weight fraction of mineral $f_i(P,T)$ to track the evolution of hydration and associated alteration assemblage. The degree of hydration $\alpha_{hyd}$ is given as the cumulated mass fraction of hydrous mineral in the rock

$$\alpha_{hyd} = \sum_i f_i(P,T). \quad (4.4)$$

At each time step $t$, for each nodal position $(k,j)$ in the model domain, we update the mineral composition and water content according to thermodynamic equilibrium. If hydration is allowed to occur at that position (based on criteria in equations 4.1 and 4.2) $f_i(P_{k,j}, T_{k,j}, t)$ and $r_{sat}^{eq}(P_{k,j}, T_{k,j}, t)$ are updated for those nodes. If hydration is not allowed at a nodal position $(k,j,t)$, then the mineral composition and water content from the previous time step are retained. This approach ensures that the hydration history and associated alteration assemblages are preserved in regions where hydration is not currently occurring, despite changes in pressure and temperature.

### 5-    Plastic and viscous weakening

Hydrous minerals have an internal coefficient of friction lower than anhydrous minerals (e.g., (5)). In nature, contributing weakening factors include foliation development, fluid-assisted processes like pressure solution and grain size reduction, and hydrous mineral formation (e.g., (53)).

In brittle regime, to account for these weakening effects, we set the mantle's friction coefficient ($\theta$) as a function of hydration degree ($\alpha_{hyd}$; equation 4.4)—which represents the cumulative abundance of hydrous minerals—and the accumulated plastic strain (I), which is the second invariant of the accumulated plastic strain.

Following (74), friction coefficient dependency on hydration degree ($\alpha_{hyd}$) is given by:

$$\theta_{hyd} = \theta_0(1 - \alpha_{hyd}) + \theta_1 \alpha_{hyd}. \quad (5.1)$$

Where $\theta_0$ and $\theta_1$ represent the friction coefficients for minimal and maximal hydration degrees, respectively.

Following (32), friction coefficient dependency on accumulated plastic strain is expressed as:

$$\theta_{acc} = \theta_0 + \frac{\theta_1 - \theta_0}{I_1 - I_0}(I - I_0), \quad (5.2)$$

where $I_0 = 0$ and $I_1 = 1$ represent the minimal and maximal values of the accumulated plastic strain, respectively, and $\theta_0$ and $\theta_1$ are the corresponding friction coefficients. The accumulated plastic strain is calculated following (*106*). The dependency on accumulated plastic strain as implemented in our model is commonly used geodynamic modelling (e.g., (*12, 14, 16, 17, 32*)).

For simplicity, we use $\theta_0 = 30° (\approx 0.6)$ and $\theta_1 = 15° (\approx 0.3)$ for the friction coefficients corresponding to both minimal and maximal values of accumulated plastic strain and hydration degrees in both equations (5.1) and (5.2).

Assuming that the overall friction coefficient ($\theta$) of the mantle is governed by the weaker of the two effects—either hydration or accumulated plastic strain—the friction coefficient is determined by taking the minimum of the two components:

$$\theta = min(\theta_{hyd}, \theta_{acc}). \qquad (5.3)$$

We define cut-off conditions for both $\theta_{acc}$ and $\theta_{hyd}$, such that

$$\begin{vmatrix} \theta_{acc} = \theta_{acc,min} = \theta_1, & for\ I \geq 1 \\ \theta_{hyd} = \theta_{acc,min}, & for\ \theta_{hyd} \leq \theta_{acc,min} \end{vmatrix} \qquad (5.4)$$

Here, $\theta_{acc,min}$ represents the minimal friction coefficient value, dependent on the accumulated plastic strain (equation 5.2). The value $\theta_{acc\_min}$ is reached when the accumulated plastic strain equals or exceeds a threshold of 1.

This constraint (equation 5.4) implies that, in high-strain fault zones, plastic strain accumulation has a more significant influence on the friction coefficient than hydration. Conversely, in region of lower strain, hydration becomes the dominant factor (see section Implications for depth-distribution of earthquakes in the main text and Fig. 7e, f).

In the ductile regime, we introduce viscous weakening to account for the effects of grain size reduction and crystallographic preferred orientation (e.g., (*53*)). We linearly increase the pre-exponential factor B of the dislocation creep law (equation 2.5) to represent the weakening mechanism due to these microstructural changes (*52, 53, 107*). Hence, the pre-exponential factor remains constant for no deformation and increases linearly with viscous deformation, up to a maximum value ($W_{max}$) times larger than the original value when the finite viscous strain reaches or exceeds 1. This maximum weakening factor ($W_{max}$) decreases assuming an Arrhenius-like relationship with temperature (T) to simulate the decreasing effect of weakening where crystal growth rates would be rapid. Specifically, $W_{max}$ is set to 30 when T ≤ 800 °C and 1 when T ≥ 1200 °C (e.g., (*14, 16*)).

SUPPLEMENTARY MATERIALS

# TECTONICS CONTROLS HYDRATION-INDUCED RHEOLOGICAL HETEROGENEITIES IN ULTRASLOW-SPREAD OCEANIC LITHOSPHERES

This supplementary contains:

Figures S1and S2
Tables S1, S2 and S3
Movies S1 to S7

Figure S1:

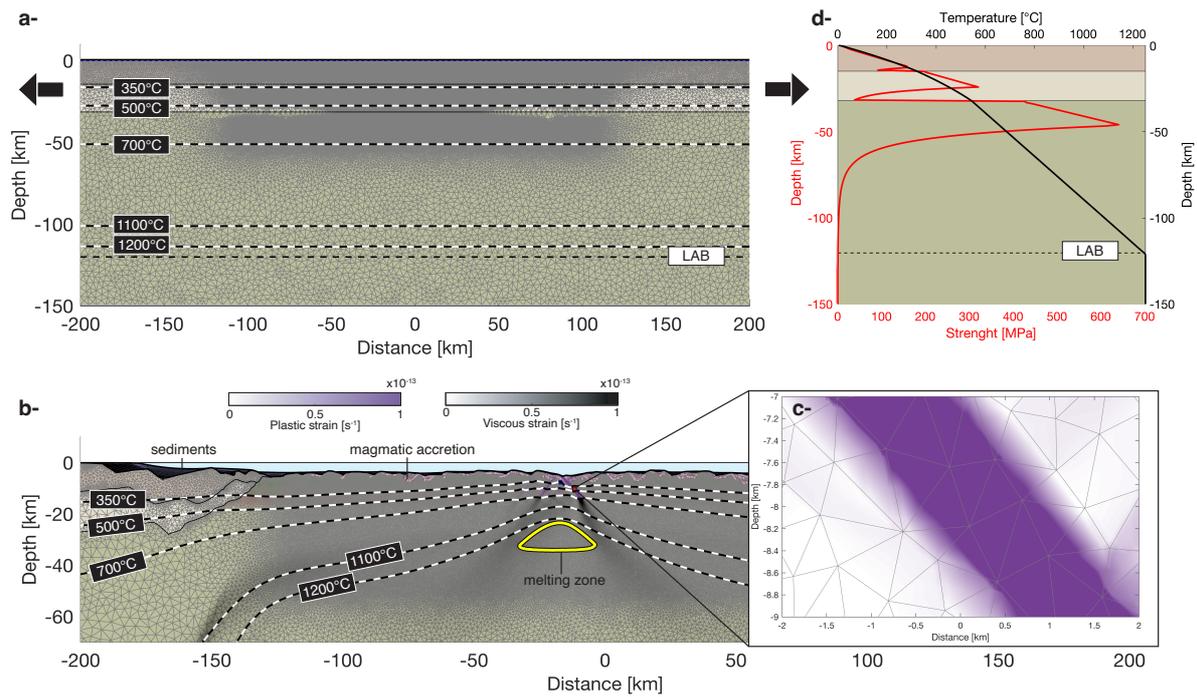

Figure S1: (a) Model setup and mesh, with isotherms shown as dashed black and white lines. The thick black arrows on either side of the model box represent the half-extension constant velocity applied on either side. The model shown here is the model for the SWIR case, which has 1250 °C for mantle temperature (set constant from the initial Lithosphere-Asthenosphere Boundary (LAB) at 120 km to the base of the model at 150 km depth) and 15 mm. yr$^{-1}$ for full spreading velocity. (b) Model development after 21.6 Myr of extension with the mesh and isotherms shown as dashed black and white lines. Sediments are shown in black. Purple markers show the accreted magma. The mantle melting zone is shown as yellow line. (c) Mesh along a segment of the detachment fault zone. Active brittle and ductile deformation are shown in purple and black transparency scales, respectively. (d) Initial strength profiles (red line) and geothermal structure (black line).

Figure S2:

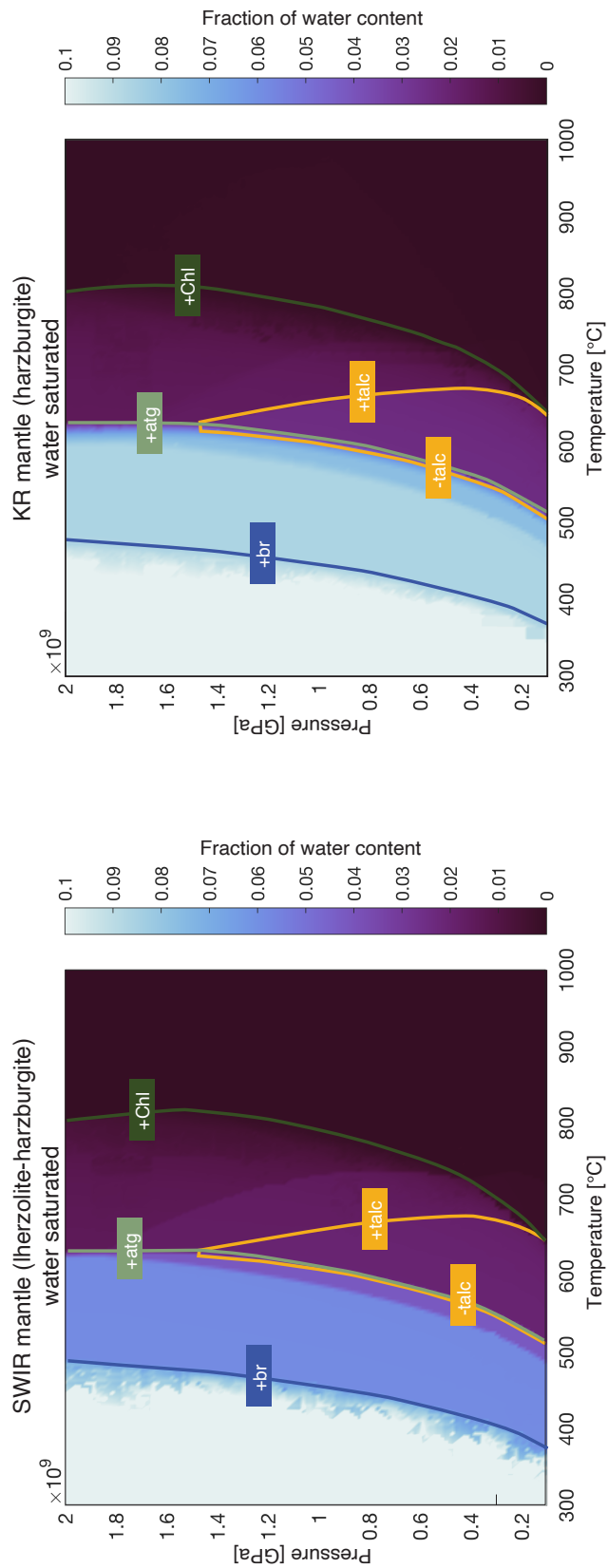

Figure S2: Fraction of water content (wt %) for SWIR (Southwest Indian Ridge) and KR (Knipovich Rigde) mantle compositions in the NCFMASH ($Na_2O$-$CaO$-$FeO$-$MgO$-$Al_2O_3$-$SiO_2$-$OH_2O$) closed system, assuming water saturation conditions. Colored lines show selected reactions (see Figure 2 in the main manuscript).

Table S1:

| Thermomechanical parameters | Wet quartzite (UC and weak LC) | Wet anorthite (intermediate strength LC) | Dry olivine (lithospheric mantle) | Wet olivine (asthenospheric mantle) |
|---|---|---|---|---|
| Dislocation pre-exponential factor $log(B_{dis})$ $[Pa^{-n}\ s^{-1}]$ | −28.0 | −15.40 | −15.96 | −15.81 |
| Dislocation exponent $n_{dis}$ | 4.0 | 3.0 | 3.5 | 3.5 |
| Dislocation activation energy $E^*_{dis}$ $[kJ\ mol^{-1}]$ | 223 | 356 | 530 | 480 |
| Dislocation activation volume $V^*_{dis}$ $[10^{-6}\ m^3\ mol^{-1}]$ | 0 | 0 | 13 | 10 |
| Diffusion pre-exponential factor $log(B_{dif})$ $[Pa^{-n}\ s^{-1}]$ | - | - | −8.16 | −8.64 |
| Diffusion exponent $n_{dif}$ | - | - | 1 | 1 |
| Diffusion activation energy $E^*_{dif}$ $[kJ\ mol^{-1}]$ | - | - | 375 | 335 |
| Diffusion activation volume $V^*_{dif}$ $[10^{-6}\ m^3\ mol^{-1}]$ | - | - | 6 | 4 |
|  | UC | LC | Asthenospheric mantle |  |
| Shear modulus $\mu$ $[GPa]$ | 36 | 40 | 74 |  |
| Thermal conductivity $k$ $[W\ m^{-1}\ K^{-1}]$ | 2.1 | 2.5 | 3.3 |  |
| Heat capacity $C_p$ $[J\ kg^{-1}\ K^{-1}]$ | 1,200 | 1,200 | 1,200 |  |
| Radiogenic heat production $H_r$ $[\mu W\ m^{-3}]$ | 1.3 | 0.73 | 0 |  |
| Reference densities $\rho_0$ $[kg\ m^{-3}]$ | 2,700 | 2,850 | 3,360 |  |
| Thermal expansivity coefficient $\alpha_T$ $[10^{-5}\ K^{-1}]$ | 2.3 | 2.4 | 3.0 |  |
| Cohesion (Initial – Final) $C_0$ $[MPa]$ | 10 - 4 |  |  |  |
| Friction angle (Initial – Final) $\phi_0$ $[°]$ | 30 - 15 |  |  |  |
| Strain weakening range | 0 - 1 |  |  |  |
| Reference density of the oceanic crust $[kg\ m^{-3}]$ | 2,750 |  |  |  |

The remaining parameters are from (*1*). Diffusion creep B is calculated using a grain size of 6 mm. Wet olivine water content is 500 ppm H/Si. Rheological parameters for upper crust (UC) (*2*), lower crust (LC), lithospheric mantle, and asthenospheric mantle are from, (*3*), and (*4*). Depletion factor for density dependence β is from (*5*).

Table S2: Normalized bulk chemical composition

| NCFMAS-Hsat | SWIR mantle | KR mantle |
|---|---|---|
| wt% | | |
| $Na_2O$ | 0.05 | 0.13 |
| CaO | 1.66 | 1.99 |
| $FeO_{tot}$ | 8.44 | 8.49 |
| MgO | 43.2 | 40.97 |
| $Al_2O_3$ | 2.11 | 2.55 |
| $SiO_2$ | 44.53 | 45.88 |

The SWIR mantle composition represents a composite harzburgite-lherzolite composition compiled from a suite mantle lherzolite and harzburgite compositions along the southwest Indian ridge (*6*). The KR mantle composition represents bulk harzburgite composition from rock samples along the sparsely magmatic zone in Gakkel Ridge (*7*, *8*).

Table S3: thermodynamic solid solution models

| Predefined solid solution model | Abbreviation | References |
|---|---|---|
| amphibole | amph | Dale et al. (2005) (5) |
| chlorite | chl | Holland and Powell (1998) (9) |
| clinopyroxene | cpx | Holland and Powell (1996) (10) |
| orthopyroxene | opx | Holland and Powell (1996) (10) |
| antigorite (serpentine) | atg | Padrón-Navarta et al. (2013) (11) |
| olivine | ol | Holland and Powell (1998) (9) |
| garnet | gt | Holland and Powell (1998) (9) |
| plagioclase | pl | Newton et al. (1980) (12) |

The thermodynamic solid solution models are the same as those reported in (13). Thermodynamic calculations were obtained using Perple_X software (14), with the following thermodynamic files hp02ver.dat and model_solution_684.dat.

**Movie captions**

Movie S1: Evolution of the accumulated plastic strain from continental rifting, continental break-up and mature seafloor spreading for the reference model, the SWIR case. For this model, the mantle temperature is set 1250 °C (constant from the initial Lithosphere-Asthenosphere Boundary (LAB) at 120 km to the base of the model) and we applied half of the full spreading rate of 15 mm. yr-1 on both side of the model. Active plastic and viscous strain are shown in purple and black transparency scales. The dashed black and white lines show the following isotherms: 350 °C, 500 °C (approximation of the serpentinisation front) and 700 °C (approximation of the deepest hydration fronts and the ductile brittle transition zone, BDT). The blue arrows represent the mantle flow pattern. Sediments are shown in black and the oceanic domain is shown in blue. The currently accreted magma is shown by the blue markers.

Movie S2: Evolution of the log10 total strain rate for the SWIR case (see caption Movie S1). The dashed black and white lines show the following isotherms: 350 °C, 500 °C (approximation of the serpentinisation front) and 700 °C (approximation of the deepest hydration fronts and the ductile brittle transition zone, BDT). The blue arrows represent the mantle flow pattern. Sediments are shown in black and the oceanic domain is shown in sky blue. The currently accreted magma is shown by the blue markers.

Movie S3: Evolution of the spatial distribution of the water for ongoing hydration along active brittle faults for the SWIR case (see caption Movie S1). The dashed black and white lines show the following isotherms: 350 °C, 500 °C (approximation of the serpentinisation front) and 700 °C (approximation of the deepest hydration fronts and the ductile brittle transition zone, BDT). Areas where hydration occurred has shown in cyan transparency. Sediments are shown in black and the oceanic domain is shown in sky blue. The currently accreted magma is shown by the blue markers.

Movie S4: Evolution of the spatial distribution of hydration in the oceanic lithosphere for the SWIR case (see caption Movie S1). Hydration is simulated using the thermodynamic predictions from the SWIR mantle composition (Supplementary Tab. S3). Active plastic and viscous strain are shown in purple and black transparency scales. The dashed black and white lines show the following isotherms: 350 °C, 500 °C (approximation of the serpentinisation front) and 700 °C (approximation of the deepest hydration fronts and the ductile brittle transition zone, BDT). Purple markers show the accreted magma. The currently accreted magma is shown by the blue markers. Sediments are shown in black and the oceanic domain is shown in sky blue. The currently accreted magma is shown by the blue markers.

Movie S5: Evolution of the spatial distribution of hydration in the oceanic lithosphere for the KR case. For this model, the mantle temperature is set 1310 °C (constant from the initial Lithosphere-Asthenosphere Boundary (LAB) at 120 km to the base of the model) and we applied half of the full spreading rate of 10 mm. yr-1 on both side of the model. Hydration is simulated using the thermodynamic predictions from the KR mantle composition (Supplementary Tab. S3). Active plastic and viscous strain are shown in purple and black transparency scales. The dashed black and white lines show the following isotherms: 350 °C, 500 °C (approximation of the serpentinisation front) and 700 °C (approximation of the deepest hydration fronts and the ductile brittle transition zone, BDT). Purple markers show the accreted magma. The currently accreted magma is shown by the blue markers. Sediments are shown in black and the oceanic domain is shown in sky blue. The currently accreted magma is shown by the blue markers.

Movie S6: Evolution of the spatial distribution of rheological change in the oceanic lithosphere for the SWIR case (see caption Movie S1). Active plastic and viscous strain are shown in purple and black transparency scales. The dashed black and white lines show the following isotherms: 350 °C, 500 °C (approximation of the serpentinisation front) and 700 °C (approximation of the deepest hydration fronts and the ductile brittle transition zone, BDT). Sediments are shown in black and the oceanic domain is shown in sky blue. The currently accreted magma is shown by the blue markers.

Movie S7: Evolution of the spatial distribution of hydration degree in the oceanic lithosphere for the SWIR case (see caption Movie S1). Active plastic and viscous strain are shown in purple and black transparency scales. The dashed black and white lines show the following isotherms: 350 °C, 500 °C (approximation of the serpentinisation front) and 700 °C (approximation of the deepest hydration fronts and the ductile brittle transition zone, BDT). The blue arrows represent the mantle flow pattern. Sediments are shown in black and the oceanic domain is shown in sky blue. The currently accreted magma is shown by the blue markers.